\documentclass[sigconf, nonacm]{acmart}

\AtBeginDocument{%
  }

\setcopyright{acmlicensed}
\copyrightyear{2025}
\acmYear{2025}
\acmDOI{XXXXXXX.XXXXXXX}

\acmConference[EICS '26]{Make sure to enter the correct
  conference title from your rights confirmation email}{June 03--05,
  2025}{Woodstock, NY}
\acmISBN{978-1-4503-XXXX-X/18/06}

\begin{document}

\title[Semantic Interactivity]{\textit{Semantic Interactivity}: leveraging NLP to enable a shared interaction approach for joint activities}






\author{Olaf V. Adan}
\email{o.v.adan@tue.nl}
\orcid{0000-0002-2133-2288}
\affiliation{
  \institution{Eindhoven University of Technology}
  \streetaddress{Den Dolech, 2}
  \city{Eindhoven}
  \country{NL}
}

\author{Dimitra Dritsa}
\email{d.dritsa@tue.nl}
\orcid{0000-0002-7615-8520}
\affiliation{
  \institution{Eindhoven University of Technology}
  \streetaddress{Den Dolech, 2}
  \city{Eindhoven}
  \country{NL}
}

\author{Steven Houben}
\email{s.houben@tue.nl}
\orcid{0000-0002-9009-5706}
\affiliation{
  \institution{Eindhoven University of Technology}
  \streetaddress{Den Dolech, 2}
  \city{Eindhoven}
  \country{NL}
}




\begin{abstract}
Collocated collaboration, where individuals work together in the same physical space and time, remains a cornerstone of effective teamwork. However, most collaborative systems are designed to support individual tasks rather than joint activities; they enable interactions for users to complete tasks rather than interactivity to engage in shared experiences. In this work, we introduce an NLP-driven mechanism that enables \textit{semantic interactivity} through a shared interaction mechanism. This mechanism was developed as part of \textit{CollEagle}, an interactive tabletop system that supports shared externalisation practices by offering a low-effort way for users to create, curate, organise, and structure information to capture the essence of collaborative discussions. Our preliminary study highlights the potential for semantic interactivity to mediate group interactions, suggesting that the interaction approach paves the way for designing novel collaborative interfaces. We contribute our implementation and offer insights for future research to enable semantic interactivity in systems that support joint activities.

\end{abstract}

\ccsdesc[500]{Human-centered computing~Interaction techniques}
\ccsdesc[300]{Human-centered computing~Natural language interfaces}
\ccsdesc[500]{Human-centered computing~Collaborative interaction}

\keywords{Collaboration, Speech Recognition, Collaborative Systems}
\begin{teaserfigure}
  \includegraphics[width=\textwidth]{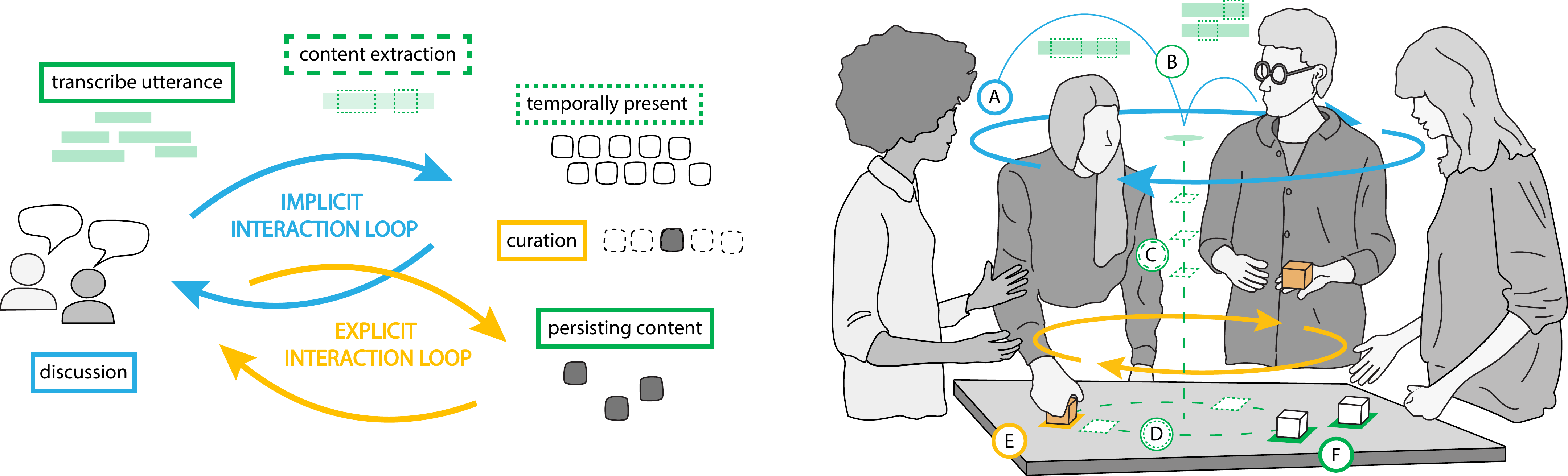}
  \caption{Left: Schematic model for the shared interaction approach, comprising an implicit (blue) and explicit (orange) interaction loop. Right: Implemented into a tangible interactive tabletop system, for which discussions (blue) are transcribed (A) per utterance (B) to extract (C) and temporally present content (green) to users (D), with tangibles (orange) enabling users to curate content (E) through collaborative action, after which content persists (F) as building material to support the joint activity.}
  \Description{The teaser figure displays how our shared interaction approach comprises two interaction loops: an implicit loop that continuously presents users with content extracted from their conversations, complemented by an explicit loop that enables users to curate and use presented content. The figure shows how the shared interaction approach operationalises an implicit loop for content creation by transcribing utterances, extracting content from these and temporally presenting these to users. In turn, this content in curated by what constitutes an explicit interaction loop, for which curating presented content causes this to persist. Complementing this figure, an image is provided showing four people engaged in collaboration around a tabletop, with the implicit loop positioned at the height of their heads and the explicit loop near their hands. This image shows the shared interaction approach implemented in CollEagle, with the implicit interaction loop defined by a stream of virtual post-its on the interface. The figure shows participants having orange blocks in their hand that are to be used for explicit interactions. In addition to the stream of virtual post-its, there are blocks on the interface that represent tangible annotations, with visualisations of the content collected resulting in these surrounding the orange blocks.}
  \label{fig:teaser}
\end{teaserfigure}


\maketitle

\section{Introduction}
Collocated collaboration, where people work together in the same physical space and time, remains an integral part of collaborative work. When remote approaches fall short, physically sharing presence enables natural multimodal communication that helps mediate shared understanding~\cite{ olson2000distance, olson2002currently, Lee2022DistractingVideoConferencingPandemic}, supporting teams to better navigate complex and ambiguous problems ~\cite{ Chen2023, jung2017}. In such settings, \textit{externalisation} -- the act of representing ideas and concepts in a shared workspace -- is key to negotiating understanding and coordinating action~\cite{langan2004mental, mathieu2000influence, reusser2015co, van2015communication, fiore2016}. Here, digital environments extend the physical workspace by offering a scalable canvases, quick access to online resources, AI-enabled automation, and remote availability of shared materials -- mitigating the need to manually curate, collate, transpose, and share outcomes post-collaboration. However, the devices typically used to access digital workspaces impact the collaborative potential of physically co-present teams~\cite{Scott2015localremotegroupawareness}.

Research investigating applications and device configurations to integrate digital workspaces with collocated collaboration is extensive, ranging from interactive whiteboards and tabletop interfaces~\cite{Reactable2007,gronbaek2020,wallace2011investigating} to multi- and cross-device systems~\cite{houben2014activityspace, Brudy2019CrossDevice}. Yet, existing and proposed technologies typically suffer from the same problems: they require substantial attention to manage, operate, and set up, also known as \textit{configuration work}~\cite{Houben2013,houben2014activityspace}, to embed them into \textit{collocated} collaborative activities -- problems that are further exacerbated in issues with managing interactions across devices~\cite{Brudy2019CrossDevice} and the need to switch between contexts and collaboration styles~\cite{Wells2020CollabAR,tang2006}. While configuration work is problematic, it is symptomatic of a deeper issue: how existing and proposed technologies fail to support joint activities through interaction. Relying on interfaces and interactivity provided by single-user devices, most collaborative systems cater to needs and activities of the individual rather than those of the group -- supporting users to \textit{perform tasks} rather than to \textit{engage in joint activities}. Consequently, externalisation remains such a demanding task that is typically assigned to a single person, inadvertently providing them with the most control over what is externalised while preventing them from fully participating in joint activities.

To address these challenges, we developed \textbf{CollEagle}, a tangible tabletop system that integrates \textit{natural language processing (NLP)} with \textit{direct manipulation} to alleviate the cognitive demand of externalisation. CollEagle automatically transcribes conversations, extracts keyphrases, and presents these as virtual post-its that can be curated and organised through tangible interaction. Through the design and deployment of CollEagle, we identified a shared interaction mechanism -- enabling ~\textbf{semantic interactivity} -- which emerged as a by-product of system development but ultimately is the most significant contribution. This \textit{semantic interaction mechanism} comprises two interwoven loops:

\begin{itemize}
    \item \textbf{Implicit loop}: the system continuously extracts content from ongoing discussions and presents it to users.
    \item \textbf{Explicit loop}: users select, curate and organise relevant content through direct manipulation.
\end{itemize}

\noindent This mechanism enables a form of \textit{mixed-initiative interaction} that does not rely on intent recognition. Instead, it relies purely on NLP-enabled filtering and enables users to take initiative by engaging with extracted materials. The resulting mixed-initiative system shapes collaborative processes by supporting collaborative discourse between humans rather than attempting to instigate a dialogue between humans and the computer. 

We conducted a preliminary study, observing four groups of 4-5 participants using CollEagle for three 8-minute rounds to gain insight on how its interaction approach influenced in-person discussions and collaborative practices. Overall, our findings suggest `curation of content' might be a more suitable interaction approach for ambiguous and complex forms of conversation-centric collocated collaboration than the classic 'create and share content'. Consolidating our insights, this paper contributes:

\begin{itemize}
    \item A novel interaction approach to enable \textit{mixed-initiative interfaces} that integrates automation and direct manipulation through implicit and explicit interaction loops
    \item A system instantiation (\textbf{CollEagle}) that operationalises this mechanism in a tangible tabletop interface.
    \item Empirical insights into how semantic interactivity mediates collaborative practices and supports externalisation.
    \item Design implications for engineering interactive systems that reduce configuration work and enhance shared activity.
\end{itemize}

\section{Background}
For digital workspaces to support collocated collaboration without creating configuration work, we must reimagine how interactivity and interface are integrated to form collaborative systems. In the following sections, we provide historical context and outline our arguments advocating interactive surfaces to prioritise joint activities in supporting collocated collaboration. 


\subsection{Collocated Collaboration}
Unique to collocated collaboration is the sharing of presence with others in physical space, enabling us to take full advantage of the intricacies embedded in how we interact as humans. Void of technology-induced delays, sharing presence in a physical space streamlines our ability to engage in grounding~\cite{clark1991grounding} and enhances it by embedding this in the shared use of physical artefacts and workspaces~\cite{martinez2019collocated}. Powered by such `\textit{collocated qualities}', in-person collaboration helps maintain a shared understanding by allowing human-human interactions to be perceived and structured through situated actions~\cite{suchman1987plans}. This natural multimodal atmosphere ~\cite{olson2002currently} allows people to have heated discussions~\cite{Lee2022DistractingVideoConferencingPandemic}, to better balance multifaceted issues~\cite{Chen2023} and to take a more converging approach to problem solving~\cite{jung2017}, emphasising the value of collocated collaboration to navigate complex and ambiguous issues. In this context, the first objective is to find a common ground; to establish a foundation that forms a baseline from which cooperative efforts can proceed ~\cite{langan2004mental, mathieu2000influence}. Providing the necessary support for this, cognitive processes, emerging through discussions in which thoughts, opinions, knowledge, and expertise are shared, are externalised into various working materials, such as annotations, sketches, and diagrams, as a means of cognitive offloading~\cite{oppl2014facilitating, fiore2016}. Such externalised materials foster group reflection, allowing those involved in collaboration to find discrepancies and reach consensus, using externalised content to negotiate, represent, and thus mediate their shared understanding~\cite{reusser2015co, van2015communication}. A common approach to this activity is to make concept or mind maps~\cite{novak2006theory,novak2010learning}, also known as affinity diagramming~\cite{kawakita1991original}, which traditionally involves the use of large sheets of paper or whiteboards and post-its~\cite{Plaue2009ConferenceRoomAsToolBox-TechAndSOcialRoutinesIncorporateMeetingSPaces}. The use of such physical materials is integral to collocated collaboration, as the \textit{collaborative action} emerging through the \textit{shared use} of these stimulates group coordination and communication~\cite{erickson2000}. However, with the increasing use of personal devices, such as tablets and laptops, to support collaboration, the way people approach working together in collocated settings has changed drastically.

\subsection{Collaborative Computing}
Understanding how computing devices support people working together in the same space, place, and time has been a central challenge in HCI since the 90s~\cite{weiser1991computer, Weiser1993}. In this context, personal devices have been instrumental in the proliferation of productivity tools and remote online collaboration, enabling novel approaches to work together, such as collaborative writing in shared documents (e.g., Google Docs \footnote{https://docs.google.com/}) and maintaining extensive concept maps through collaborative platforms (e.g., Miro\footnote{https://miro.com/}). Evidently, the use of personal devices to access a shared canvas as a workspace for collaboration offers advantages over the use of physically situated surfaces and materials. Although, in general, personal devices are extremely useful for supporting collaboration, the consequences to their use for supporting collaboration in collocated settings must be addressed. For example, their use unnecessarily distributes work activities across displays, creating \textit{configuration work}~\cite{Houben2013,houben2014activityspace} to effectively navigate and coordinate collaborative activities. Moreover, the way personal devices enable access to digital workspaces, and corresponding interactivity, inadvertently omits the shared use of physical artefacts in collaboration, negating the emergence of collaborative actions that support group coordination. In this context, Scott et al.~\cite{Scott2015localremotegroupawareness} suggested \textit{``to incorporate interface concepts commonly used in remote collaboration to ‘add back’ awareness features that often ‘come for free’ in a co-located workspace''}. However, if anything, their remark emphasises why personal devices are poorly formatted to support collocated collaboration.

\section{Related Work}
To understand how to better integrate externalisation methods with collaborative actions, we examine current interaction techniques for externalisation in digital workspaces. Here, we find that the status quo for externalisation, in which content is \textit{created individually}, causes this to emerge as a \textit{secondary task} and, thus, fragments joint collocated work activities.

\subsection{Externalisation techniques}
With the ubiquity of personal devices, research on collaborative systems shifted from interactive whiteboards and tabletop interfaces~\cite{Reactable2007,gronbaek2020,wallace2011investigating} to multi- and cross-device systems~\cite{houben2014activityspace, Brudy2019CrossDevice}. Capturing the breadth of research efforts investigating technologies to support collocated collaborative work, Brudy et al.~\cite{Brudy2019CrossDevice} contributed a taxonomy for cross-device interactions. Examining their comprehensive overviews detailing interaction techniques, we find that multi- and cross-device interactivity often comprises techniques that aim to alleviate configuration work; supporting interactions to \textit{manage}, \textit{share}, \textit{manipulate} and \textit{curate} collaborative material~\cite{Brudy2019CrossDevice}. Notably, research investigating interactions for the creation of such materials appears to be limited, particularly with regard to interaction techniques for enabling text input. One explanation for this deficit, when it comes to multi-device setups, is that novel techniques for text input are overlooked because keyboards are already embedded in mobile devices, making the exploration of novel techniques a futile endeavour. However, for research investigating tabletop systems and interactive whiteboards as a single shared interface, physical~\cite{Clayphan2011Firestorm, Clayphan2014scriptstorm} or virtual keyboards~\cite{Martinez2011Collaid, Klinkhammer2018tabletop} are typically provided to enable text input. Beyond these, other approaches to enable text input for shared interfaces allow for writing by hand, as supported in the `Tabletop Concept Mapping' tool~\cite{oppl2009conceptmaptabletop}, or enable text input by speaking into a microphone -- as seen in `Wordplay'~\cite{hunter2008wordplay}. Although such text input techniques appear distinctly different from the use of a keyboard, users must still actively engage in, and attend to, a secondary task for text to appear. That is, while these replace the \textit{input device} and, thus, the action required to input text, users must still engage in an action that requires them to shift their attention away from joint activities.


 
\subsection{Intent in Externalisation}
Resulting from the \textit{approach behind} interaction techniques for text input, externalisations, upon creation, are inherently subject to the understanding and perspective of the person that brings the representation of subject matter into existence -- including their \textit{ intention}. Consequently, externalised materials possibly deviate from the mutually shared perspective and knowledge of the group. This discrepancy stimulates collaborative discourse, through which externalisations are revised and refined ~\cite{fiore2010,fiore2016}, and is typically an argument to spend time generating such materials; individually producing post-its to stimulate convergence in brainstorm sessions is a common approach~\cite{barki2001small}. However, it is once a group \textit{engages with} externalisations, rather than \textit{in creating} these, that collaborative discourse is stimulated~\cite{al2018review}. That is, while we must consider how externalisation techniques serve \textit{and} affect the specific purpose of collaborating in a collocated setting. For example, to optimise individual information retention and learning outcomes, handwriting is a more effective externalisation approach~\cite{mueller2014c}, specifically \textit{because} writing by hand is more cognitively demanding ~\cite{divesta1973}. Considering that the ability to engage in debate and discussion on complex matters is what makes collocated collaboration a preferred way of working together, it is these that technologies should aim to support -- to quote Erickson \& Kellogg~\cite{erickson2000}: \newline

\begin{quote}
    \textit{``It is through conversation that we create, develop, validate, and share knowledge.''} \newline
\end{quote} 

\noindent When it comes to collocated collaboration, time could be spent more effectively \textit{discussing} rather than \textit{ externalising} -- however, both are necessary for users to engage with externalisations in ways to allow for constructive negotiation of their understanding of complex issues.


\subsection{Automating Externalisation}
Providing concrete examples of how cognitive efforts for externalisation can be completely mitigated, previous works ~\cite{chandra1029TalkTracesRealtimeCaptureandVisofVerbalContentinMeetings,Shi2017talktrawallCreativeCollabVisualStimuli,Bergstrom2009} provide externalisation support by using natural language processing (NLP) to visualise recently discussed topics. Contrary to common text input techniques, these implement speech recognition as a continuous input source, intermittently processing transcripts using NLP, providing externalisations without the need for users to engage in a secondary task. However, in fully automating the externalisation process, ways for users to engage with externalisations remain limited. That is, TalkTraces~\cite{chandra1029TalkTracesRealtimeCaptureandVisofVerbalContentinMeetings} limits user interaction to inspecting and evaluating meeting transcripts, topics, keywords and statistics underlying these, and IdeaWall~\cite{Shi2017talktrawallCreativeCollabVisualStimuli}, dictating which 21 keywords remain on the interface based on a word co-occurrence graph, only enables users to cycle through several images paired to each keyword. In contrast, Conversation Clusters~\cite{Bergstrom2009} also provide users with means to mediate visualised content, supporting interactions to \textit{move} and \textit{remove} recently discussed topics. However, because the topics presented automatically disappear once the conversation moves on, the use of this content to support externalisation practices in collaboration remains beyond reach: no interaction techniques are provided for externalised materials to persist. Interactivity to mediate persistence allows collaborative use of extracted topics for externalisation, providing users with building materials to create reference points for ongoing collaboration and allowing them to construct a starting point for follow-up collaborations~\cite{bardram2005activity,brudy2016curationspace}.

\subsection{Summary}
Collocated collaboration helps navigate complex issues by enabling natural human-human communication. However, externalisation is required to effectively mediate and negotiate shared understanding, for which digital workspaces provide advantages by integrating with the broader scope of collaborative processes. Resulting from the common approach by which text input is enabled for creating externalisations in digital workspaces, collaborative systems primarily support the execution of individual tasks. Therefore, our aim is to enable shared interaction approaches for externalisation in collocated settings. In this context, interaction techniques that mediate the persistence of automatically created externalisations will allow users to engage with automatically produced externalisations -- enabling the collaborative use of these as building materials to represent and negotiate their shared understanding.


\section{Designing Collocated Interfaces}
Our aim is to design shared interfaces that accommodate to the dynamic multimodal nature of collocated collaboration, integrating these with the unique quality of sharing presence, and supporting these with computational conveniences. We aim to expand on previous work, using NLP to visualise discussions (e.g. ~\cite{chandra1029TalkTracesRealtimeCaptureandVisofVerbalContentinMeetings,Shi2017talktrawallCreativeCollabVisualStimuli,Bergstrom2009}), enabling interactions to mediate the persistence of visualised topics, turning these into building materials to collaboratively construct externalisations. While this approach can alleviate the cognitive demand of externalisation, it does not yet integrate with the unique quality of collocated settings through the  use of physical artefacts.

\subsection{Guiding principles}
The potential of tangible interactive systems to take advantage of the rich interactive qualities of physical materials has long been emphasised~\cite{price2004letsgetohys,holmquist1999token}. In this context, Oppl \& Stary~\cite{oppl2014facilitating} suggested that the integration of externalisation methods and social interaction is the key for technologies to effectively support collocated collaboration. Previous works have implemented tabletops to stimulate conversation and discussion by leveraging physically shared artefacts (e.g., ~\cite{jaasma2017,thompson2021ambidots}), including approaches for users to tangibly interact with externalised thoughts and ideas ~\cite{oppl2009conceptmaptabletop,oppl2014facilitating}. Inspired by these, we set out to implement the envisioned NLP-driven interactivity into a tangible interactive tabletop system.
Further guiding our design approach, we construed four guiding principles to designing a collocated interface that combine our arguments with insights from previous work:

\begin{enumerate}
     \item[D1] \textbf{Mitigate configuration work}  -- 
       Mitigate configuration work~\cite{Houben2016} by ensuring that interaction techniques always serve the group and collaborative efforts as a whole, allowing collaborative materials to be mediated by shared efforts, preventing switches between context and collaboration styles ~\cite{Wells2020CollabAR,tang2006}.
     \item[D2] \textbf{Balance automation and engagement} --
       Automate the creation of externalised materials while leaving it up to the users whether and how to engage with materials to support their collaborative activities; ensure automations enable continuous negotiation and representation of grounded concepts~\cite{reusser2015co, van2015communication}.
    \item[D3] \textbf{Prioritise and foster joint activities} -- 
       Complex problem solving in collaboration involves closely coupled work \cite{olson2000distance}. Keep users engaged with each other in shared tasks, limiting interactions to a turn-based format to stimulate discussions~\cite{morris2006}.  Design interfaces to be perceived as shared workspaces, ensuring that these are perceivable from all possible angles to ensure a shared focus of attention~\cite{Scott2015localremotegroupawareness}
     \item[D4] \textbf{Merge physical and digital means} --
       Leverage and enhance the qualities of collocated collaboration by ensuring that digital workspaces are accessed using physically shared materials~\cite{oppl2009conceptmaptabletop,oppl2014facilitating,Jensen2018digitizingtools}, allowing for interactivity to emerge that supports group coordination and communication ~\cite{jaasma2017,thompson2021ambidots,erickson2000}.
\end{enumerate}

\begin{figure}[h]
   \centering
   \includegraphics[width=1.0\linewidth, trim=20 25 30 70, clip]{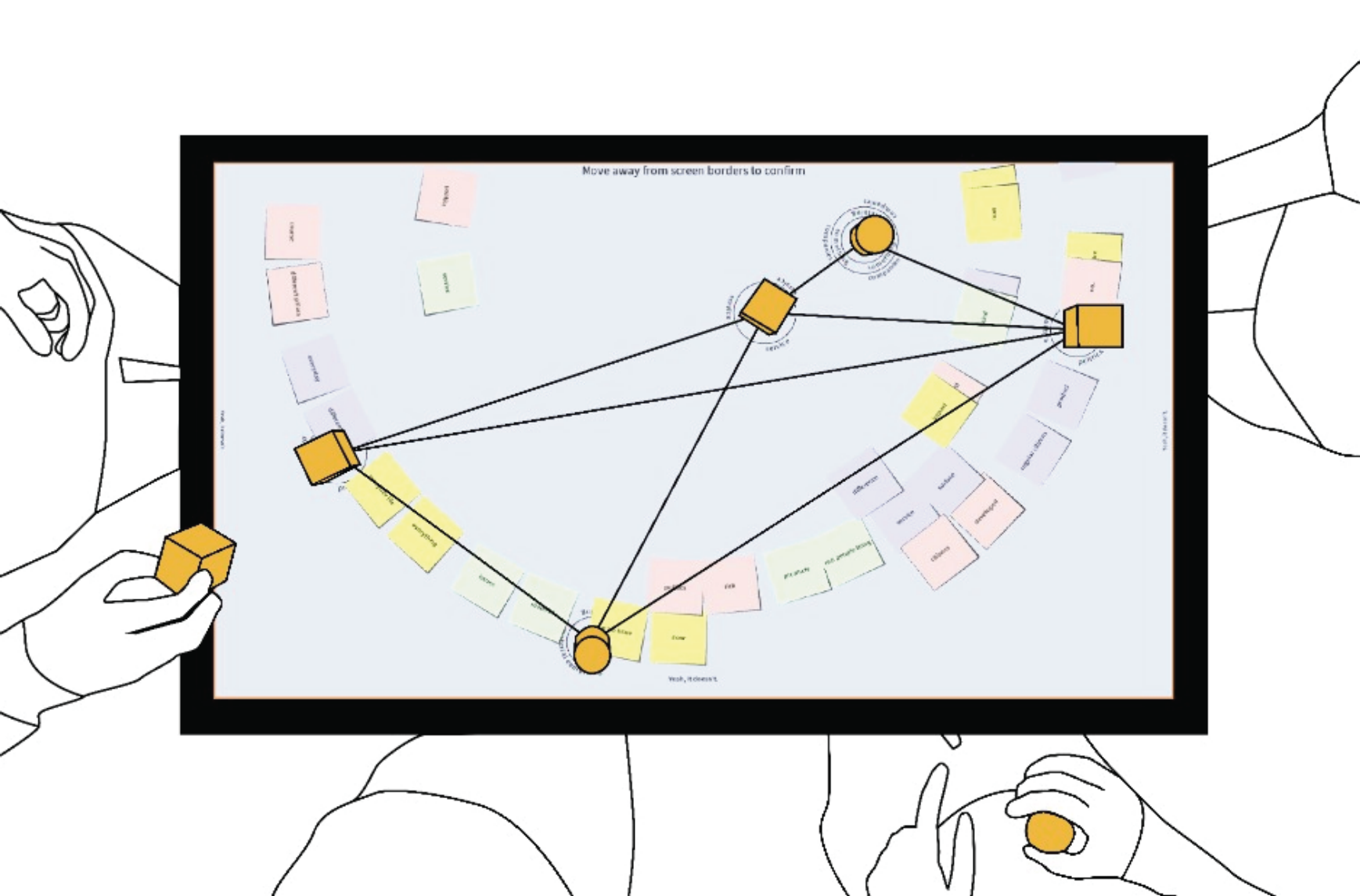}
   \caption{The CollEagle system supports the use of tangibles (orange) to interact with automatically produced post-its.}
   \Description{The figure comprises a view of participants using the CollEagle system from above. It shows the shared interaction approach implemented in CollEagle, with the implicit interaction loop actuating a stream of virtual post-its on the interface. The figure shows participants having orange blocks in their hand that are to be used for explicit interactions. In addition to the stream of virtual post-its, there are blocks on the interface that represent tangible annotations, with visualisations of the content collected resulting in these surrounding the orange blocks.}
    \label{fig:coll}
 \end{figure}

\subsection{CollEagle}
 We constructed \textit{CollEagle}~\cite{adan2023a}: a \textit{tangible interactive tabletop system} \textbf{(D1, D4)} that enables an externalisation practice \textit{without the need for users to engage in secondary tasks} \textbf{(D2, D3)}. CollEagle continuously extracts keywords from the ongoing discussion to provide users with a continuous stream of externalisations \textbf{(D2)}. Displayed as virtual post-it notes (see Figure \ref{fig:coll}), externalisations rotate in and out of the display \textbf{(D3)}, for which users can select those relevant to support their activities using tangible artefacts \textbf{(D2)}. Selecting an externalisation combines artefact and virtual post-it into a tangible annotation, enabling users to organise externalisations to create an overview of their discussions using physically shared materials \textbf{(D4)}.

 
%
 
  \begin{figure*}
   \centering
   \includegraphics[width=1.0\linewidth,trim=0 100 0 0, clip]{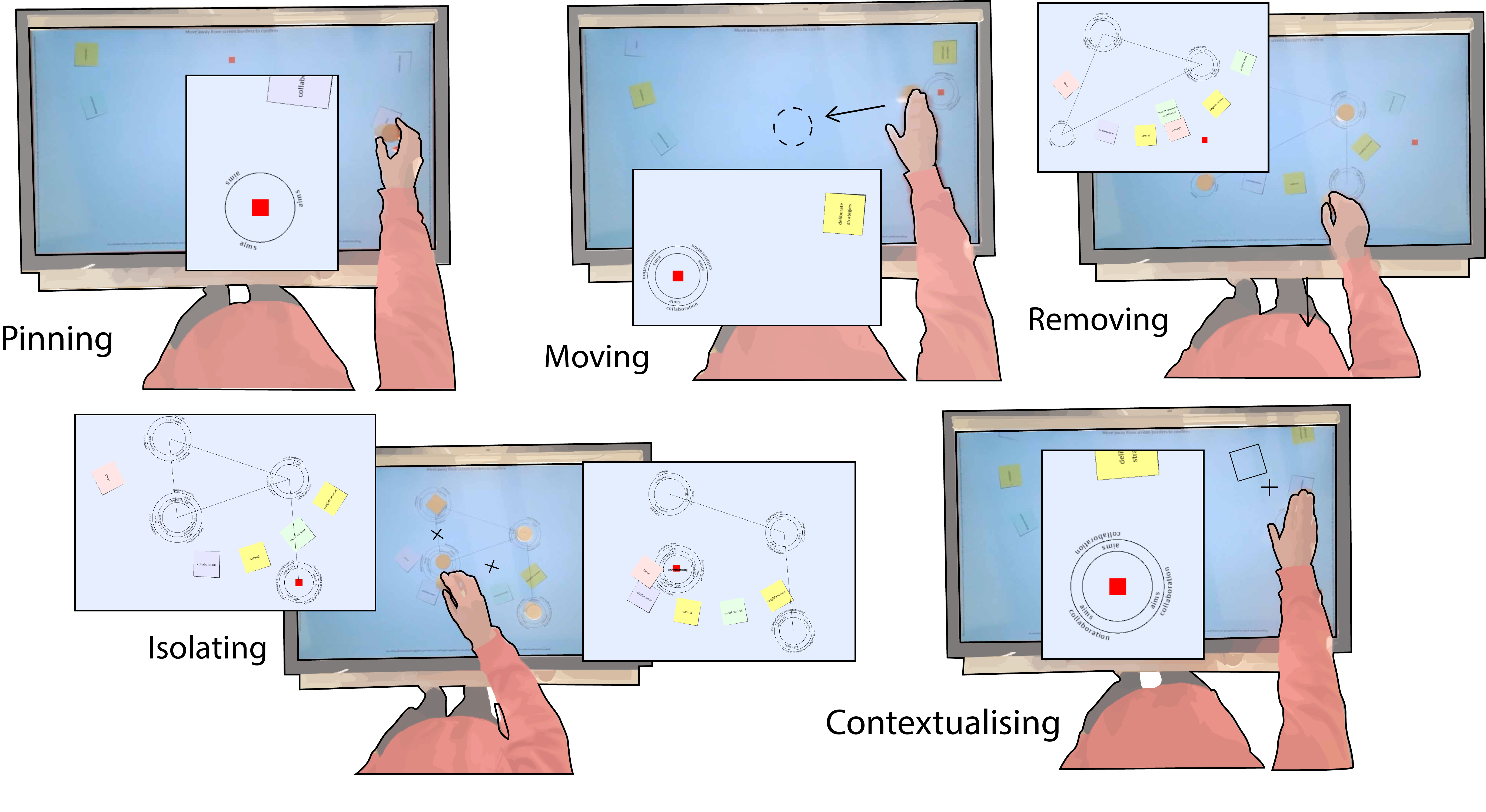}
   \caption{Digital components to tangible annotations for each interaction technique supported by CollEagle}
   \Description{This figure shows the interactions supported by CollEagle, with each supported with images to denote their effect on content on the interface. How these work is described in the text.}
    \label{fig:interact-type}
 \end{figure*}
 
\subsubsection{Interface Design}
Informing users about the origins of post-its presented in the stream, each side of the interface displays the last recognised utterances \textbf{(D1, D4}). Post-its move in and out from the top of the display in a circular fashion, remaining oriented to the nearest side of the display to provide users with equal opportunity to inspect and use them \textbf{(D1}, see Figure \ref{fig:coll}). Consecutively extracted collections can enter the display by alternating between two of four paths, with each set of paths moving post-its in the opposite directions, preventing these from overlapping (see Figure \ref{fig:interface}). Upon selection, a post-it transforms to augment the artefact used for interacting, after which the content it contained encircles the artefact used, and rotates around it to ensure readability from all possible directions (see Figures \ref{fig:coll} \& \ref{fig:interact-type}). Supporting users to identify relating concepts, CollEagle automatically draws lines between tangible annotations when the materials collected by these were extracted from the same utterance. 

\subsubsection{Interaction Techniques} 
Interacting with CollEagle provides users with tangible annotations, enabling them to \textit{mediate persistence of automatically produced materials through collaborative actions} ~\textbf{(D2,D4)}. Supported interactions are specified to cover the most basic imaginable usage of artefacts: \textit{placing}, \textit{moving} and \textit{removing}. Once created, a tangible annotation, in turn, can be moved to another location on the interface to enable spatial reasoning or removed from the interface to decouple artefact and post-its. Limiting interaction to a turn-based format, CollEagle enforces a shared interaction approach, stimulating collaborative action and discussion to ensure the collaboration remains a joint activity ~\textbf{(D1, D2, D3)}.

\subsubsection{Leveraging Modalities}
Tangible annotations, comprising digital and physical components \textbf{(D3)}, enable approaches for digitally enhancing interactivity in physically engaging with externalisations and vice versa \textbf{(D1)}.  CollEagle exemplifies this by allowing tangible annotations to collect additional materials, enabling users to \textit{contextualise} information construct externalisations in which an artefact represents a concept that unifies the layers of information augmented by its virtual counterparts (\textbf{D3, D4}, see Figure \ref{fig:coll}). With tangible interactivity implemented using a depth sensor, any opaque item can be used to interact with CollEagle -- digital components are automatically scaled to fit the dimensions of the artefact. Additionally, this implementation allows for CollEagle support interactivity by placing an artefact on top of a tangible annotation, allowing users to delete automatically drawn lines and \textit{isolate} a tangible annotation from others. \newline

\noindent Figure \ref{fig:interact-type} provides a visual overview of the interactions supported by CollEagle, designed to enable collaborative content creation through collaborative action, as described below: 
 
\begin{description} 
 \item[Pinning --] A virtual post-it is pinned into place by placing any opaque artefact on top of it, coupling the artefact and post-it, turning these into a tangible annotation.
 \item[Moving --] Tangible annotations can be moved to a different location on the interface, enabling users to arrange these and create externalisations.
\item[Removing --] Removing an artefact from the display disbands the tangible annotation, returning its digital counterpart into the stream as a virtual post-it.
\item[Isolating --] Users can stack another artefact on top of tangible annotations to isolate it from tangible annotations automatically linked by the system and delete the lines drawn between them.
\item[Contextualising --] Tangible annotations can pin additional post-its into place, enabling users to enhance the meaning represented by the artefact incrementally. 
\end{description}


 \subsection{Technical Implementation} 
\begin{figure}[h]
  \centering
  \includegraphics[width=0.75\linewidth, trim=0 26 0 71, clip]{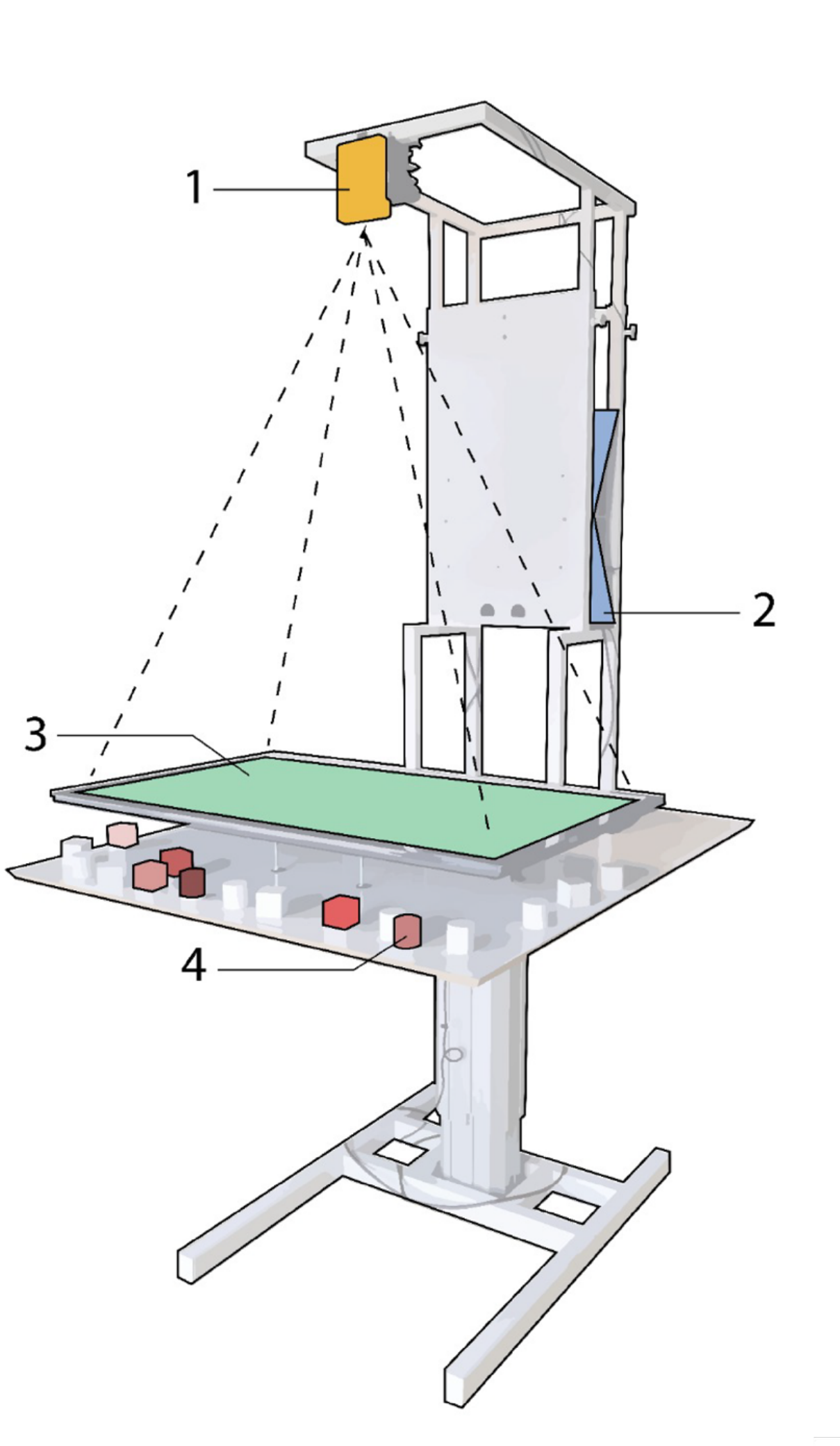}
  \caption{CollEagle System Prototype: (1) Azure Kinect Depth Sensor (2) Laptop (3) LCD-display (4) Tangibles}
  \Description{This figure shows the technical implementation of the CollEagle System, consisting of an (1) Azure Kinect Depth Sensor (2) Laptop (3) LCD-display (4) and Tangibles.}
    \label{fig:appa}
\end{figure}

The physical setup (see Figure ~\ref{fig:appa}) consists of a computer with 64GB of RAM and 6GB of VRAM, an Azure Kinect Sensor\footnote{https://azure.microsoft.com/en-us/products/kinect-dk}, and a 42” LCD screen, supplemented with 16 tangible artefacts (eight cubes and eight cylinders) made from 3mm foam-core cardboard. The Azure Kinect Sensor is attached to a metal frame and aims down at the monitor at a distance of approximately 1m. The software architecture (see Figure ~\ref{fig:softarch}) comprises i) a graphical multi-user interface built in Processing 4\footnote{https://processing.org/}, ii) a locally hosted server built using React.js\footnote{https://react.dev/} and Express.js\footnote{http://expressjs.com/} to handle NLP-services and a iii) console application built using the Platform for Situated Intelligence~\cite{bohus2021platformPSi} is used for the depth sensor to detect interactions.

 \subsubsection{Extracting content}
\begin{figure}
  \centering
  \includegraphics[width=1.0\linewidth]{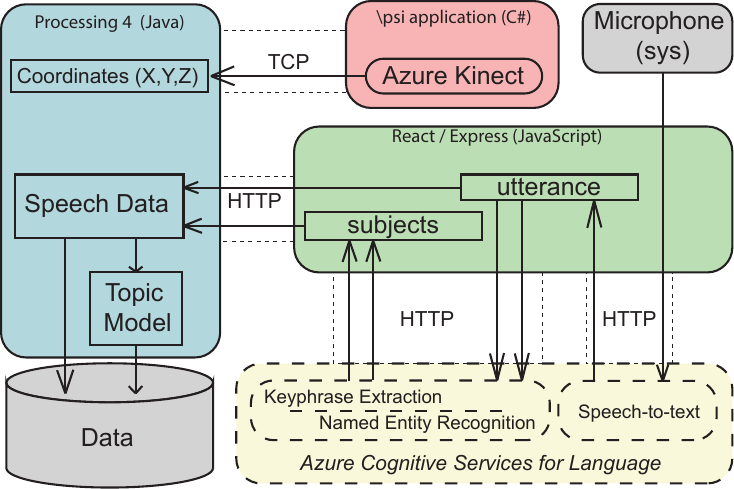}
  \caption{Software architecture.}
  \Description{This figure shows how the backend of the interface is designed, specifically regarding content extraction and the presentation of this, detailing an identified utterance with boxes encapsulating extracted contents, the four paths of the stream, how an extracted subject is displayed as post it, and where an utterance from which content originates it presented to users. The figure also includes a schematic overview that provides details of the system software architecture.}
    \label{fig:softarch}
\end{figure}

\begin{figure}
  \centering
  \includegraphics[width=1.0\linewidth]{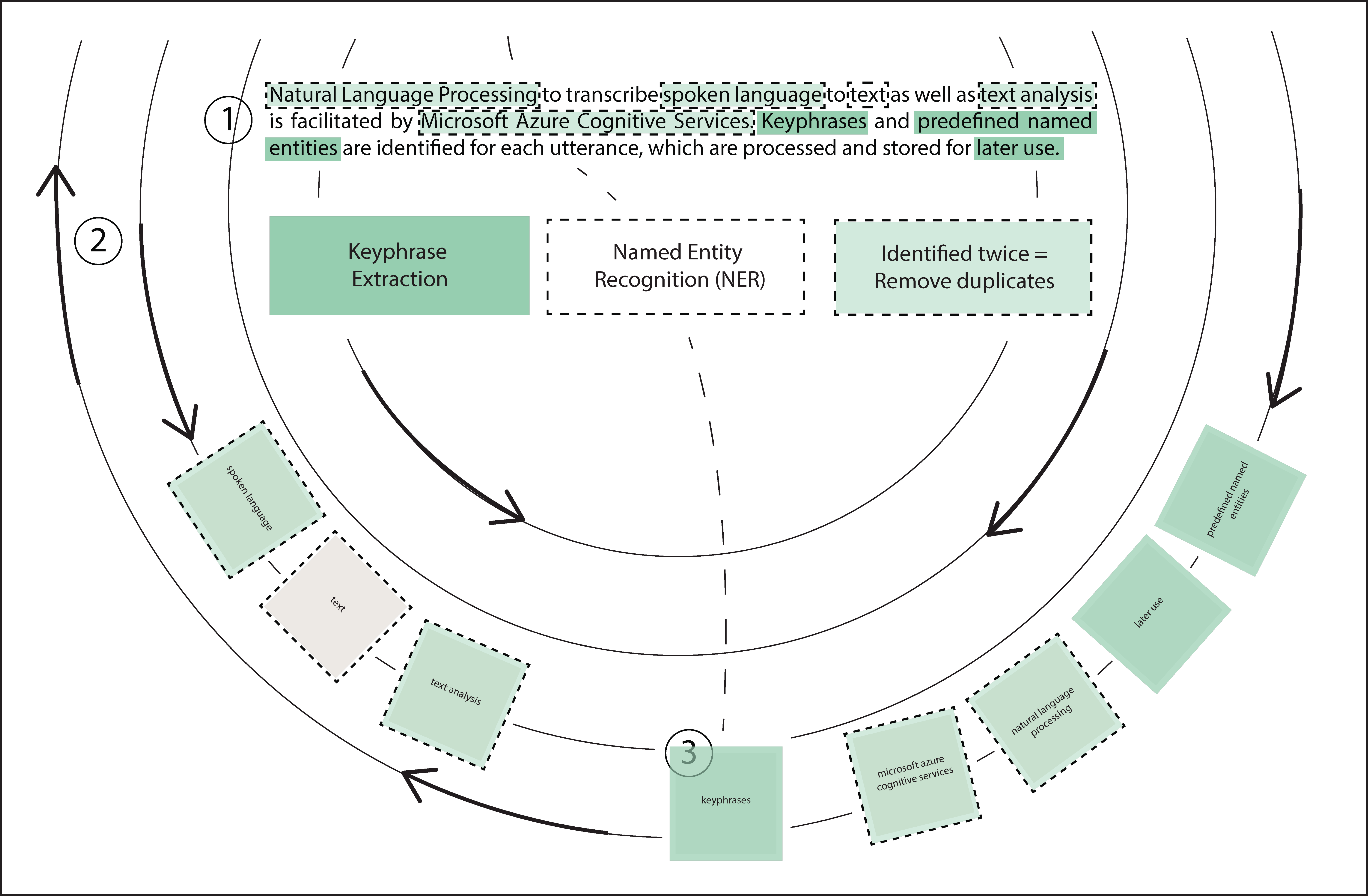}
  \caption{Schematic overview of interface design and content extraction, detailing how an (1) identified utterance is filtered to identify content, (2) stream paths alternate, and (3) content is displayed.}
  \Description{This figure shows how the backend of the interface is designed, specifically regarding content extraction and the presentation of this, detailing an identified utterance with boxes encapsulating extracted contents, the four paths of the stream, how an extracted subject is displayed as post it, and where an utterance from which content originates it presented to users. The figure also includes a schematic overview that provides details of the system software architecture.}
    \label{fig:interface}
\end{figure}

The content stream is created using the NLP and NLU modules provided by Azure Cognitive Services for Language~\cite{AzureCogn}. The ``speech-to-text'' module enables utterances up to 15 seconds to be transcribed, which, given conversation continues, means content (i.e., a transcribed utterance) is produced continuously after at most 15 seconds. Using this to produce bite-sized content, incoming utterances are immediately processed using ``keyphrase extraction'' and ``Named Entity Recognition'' to produce a set of key subjects. (see Figure \ref{fig:softarch}).

\subsubsection{Topic Modelling} 
Supporting users with insight into relationships between discussed key subjects during their conversation, CollEagle implements a simple bag-of-words algorithm to draw lines between subjects discussed in relation to each other. This algorithm relies on specifications of the ``speech-to-text'' service, by which each recognised utterance is taken as a unit of information that captures conceptual understanding. With the stream implemented to continuously process and extract content using NLP and NLU, this unit of information, and corresponding set of key subjects, enables the creation of a dataset consisting of (i) identified utterances and subjects that are extracted from these, (ii) the number of times subjects were identified, and (iii) a topic model describing which subjects have been extracted in parallel from utterances and how often these occurred. 

 \subsubsection{Depth-map actuated interaction} 
An Azure Kinect sensor is implemented to create a 3D interaction space and support token-agnostic interaction. The centre of an artefact is used as a pointer, and physical dimensions are stored to ensure visual components to tangible annotations correspond the size of artefacts. The system continuously compares stored- to incoming depth maps unless users cross the border of the interaction space. Preventing users to interact simultaneously, interactivity is operationalised by identifying the outer bounds of height changes, for which the system stores a depth map at startup and another to every time an interaction is completed as users out of the interaction space. Depth pixel coordinates with increased and decreased height values are used to determine whether an artefact was placed or removed, a combination of these enables the system to identify where an item moved on the interface. Accounting for changes between depth- and display pixel density, registered coordinates are processed relative to the height of the artefact used in interaction; the sensor is positioned to define a box of depth pixels aligned with the corners of the interface, for which each pixel holds information regarding their distance from a plane perpendicular to the forward orientation of the sensor (see Figure \ref{fig:appa}).


\section{Semantic Interactivity}
\begin{figure}[b]
   \centering
   \includegraphics[width=1.0\linewidth]{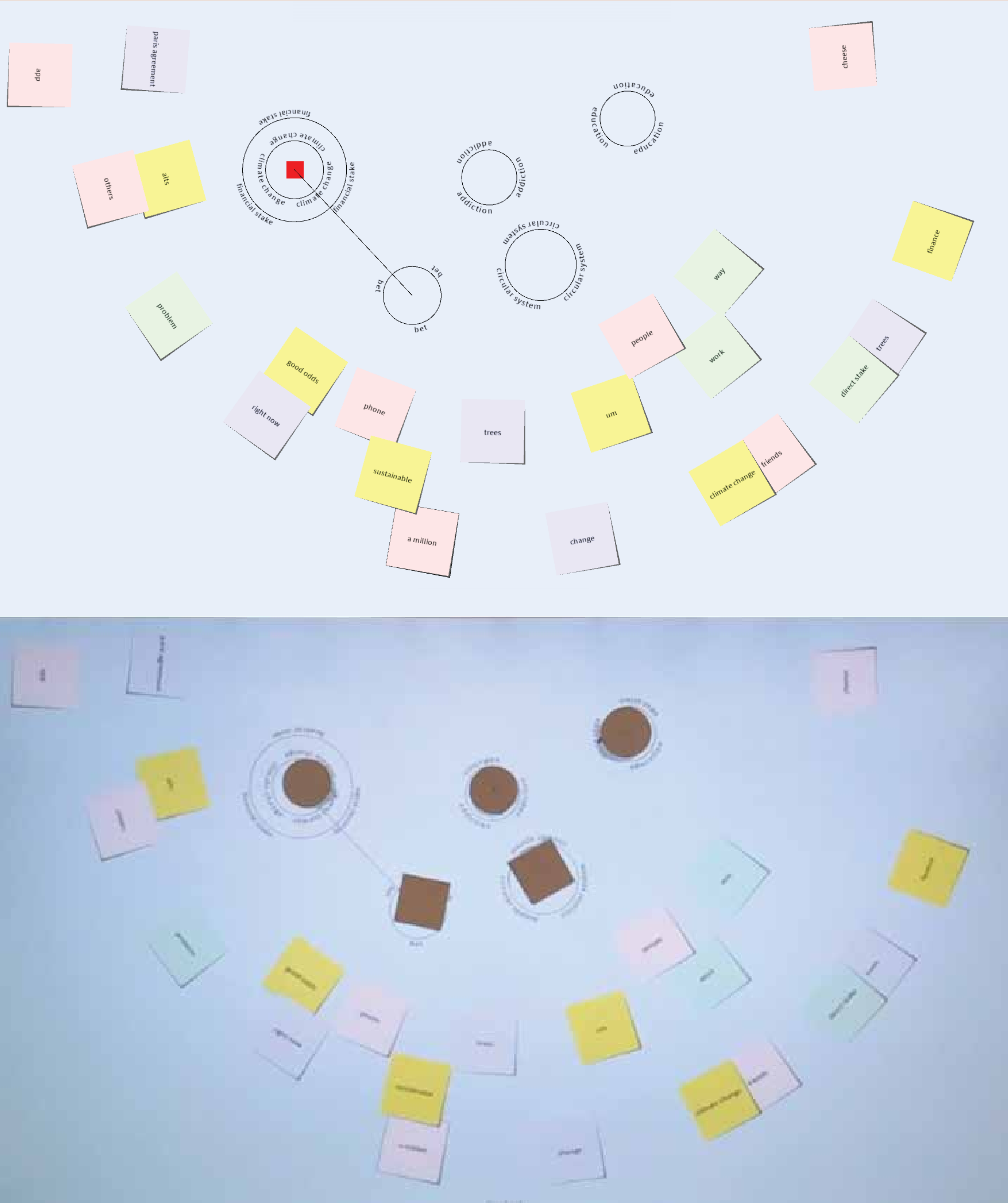}
   \caption{Two loops emerged in automatically producing content (digital) and tangibly manipulating it (physical).}
   \Description{This figure consists of two images, one being a screenshot of the interface that shows virtual components to tangible annotations, the other is a photo of the interface with the same screen to show the artefacts on the display as tangible annotations.}
    \label{fig:physdig}
 \end{figure}

The purpose of CollEagle is to continuously provide users with building materials to support collaborative activities on a shared interface, for which it (i) automatically produces materials that reflect the discussion ~\cite{Shi2017talktrawallCreativeCollabVisualStimuli,chandra1029TalkTracesRealtimeCaptureandVisofVerbalContentinMeetings,Bergstrom2009} and (ii) enables direct manipulation~\cite{shneiderman1983direct,hutchins1985direct} of these through tangible interactions. Resulting from the integration of its two interactive loops, situated in the digital and physical (see Figure \ref{fig:physdig}), CollEagle supports joint activities in collocated work in a way that makes visible a distinct approach to enable mixed-initiative interactivity~\cite{allen1999mixed}. 

\subsection{Mixed-Initiative Interaction}
Horvitz ~\cite{horvitz1999uncertainty} emphasised the potential for mixed-initiative interaction to support joint activities in the late 90s, particularly for systems to integrate effectively with the dynamic nature of collaborative discussion. In this interaction approach, user- and system-initiated actions are envisioned to complement each other in performing collaborative tasks in a way that resembles how humans engage in joint problem solving~\cite{allen1999mixed}. Unique to CollEagle's approach is that the task of creating externalisations is actuated without the need the system to infer the user's intent. This suggests that CollEagle's implementation lies in the \textit{ fixed-subtask level} of mixed-initiative interaction. On the other hand, the system continuously monitors the conversation and coordinates its subtask of extracting and producing content based on the underlying language model, suggesting that it operates at the \textit{ negotiated level}. 

\subsection{Shared Interaction Approach}
CollEagle's approach to mixed-initiative interaction is defined by \textit{the extent to which interactivity is mediated by user intent in supporting joint activities}. Making this concrete, we propose a \textit{semantic interaction mechanism} (see Figure \ref{fig:teaser}) enables mixed-initiative interfaces that \textit{automate material production without relying on user intent, for which direct manipulation enables these materials to persist}. Specifically, with the creation of collaborative materials occurring as a side-effect of humans interacting with one another, we suggest that the interaction loops that enable \textbf{semantic interactivity} are best defined by building on Serim \& Jacucci's~\cite{serim2019} intentionality-based definition to distinguish implicit and explicit interaction:




\subsubsection*{\textbf{Implicit interaction}} 
The implicit interaction loop monitors collaborative discussion to continuously extracting key subjects and producing content into bite-sized pieces for these to be perceived as building materials. These are temporarily presented to users to prevent materials from overpopulating the interface while ensuring that only materials deemed valuable by the user persist on the interface. Effective implementations for the implicit loop provide users with enough time to evaluate their value before disappearing.

\subsubsection*{\textbf{Explicit interaction}}
The explicit interaction loop enables the direct manipulation of building materials in support of collaborative activities. At its core, it comprises interactions that \textit{enable the use} of implicitly created materials and support users to leverage what they deem appropriate to navigate, mediate, and support joint activities. At the most basic level, it enables interactions for presented materials to persist, be positioned, and be deleted. \newline

\noindent We ascribe the term ``semantic'' to the proposed mechanism to emphasise that both interaction loops are mediated by natural language, providing users ways to draw meaning from these, supporting collaboration by enabling explicit, implicit, and -- combining these -- semantic interactivity.


\section{Study}
We conducted an observational study to get preliminary insights into how participants used CollEagle to externalise and support their discussions. Specifically, we examined (i) how \textit{ interactions} with the system \textit{ influence collaboration}, (ii) how automatically produced \textit{externalisations}, sourced from ongoing conversation, \textit{ impact} the trajectory and flow of \textit{discussions} and (iii) the \textit{perceived value} of the externalisation practice, interface and interactions provided by the system. 

\subsection*{\fontsize{9}{12}\selectfont Participants}
We recruited 17 participants between the ages of 20-31 through convenience sampling, resulting in four groups. One group (G1, n = 5) comprised the board of an interdisciplinary undergraduate student team, two of graduate students (G2, G3 n = 4) and the last of PhD researchers (G4, n = 4). None of the participants were native English speakers.

\subsection*{\fontsize{9}{12}\selectfont  Procedure}
The interaction approach of the system was introduced to the participants through a live demonstration. The participants were explained how the content stream was produced from their discussions and shown how to curate material from the stream using the tangibles. After each supported interaction (pinning, moving, stacking, contextualising \& removing) was demonstrated, participants were given some time to familiarise themselves with these. The study started after every participant at least successfully created a tangible annotation, moved it to another place, and removed it from the interface. 

\subsection*{\fontsize{9}{12}\selectfont Task}
Participants were instructed to make an overview that captured their collaborative efforts during three 8-minute rounds. Specifically, participants were asked to (i) ``use the system and make an overview of your discussion'',  (ii) ``use the system to conceptualise ideas based on your discussion'', and (iii) ``use the system to make a plan to implement your ideas''. Sessions were guided by the question: ``How to \textit{really} solve climate change?'' to provide groups with a topic for their collaborative endeavour. Each of the groups discussed this societal challenge during the first round, after which this continued to guide the following rounds. An exception to this this was G1, which we allowed to use the second and third rounds, conceptualising ideas and making a plan, respectively, to conduct a scheduled board meeting to organise an event, providing us with the opportunity to observe how these used the system in a non-simulated real-world context.

\subsection*{\fontsize{9}{12}\selectfont Finalisation}
Concluding each study, we conducted semi-structured interviews, formatted as group discussions, to gain insight into user perspectives relating to their experience and perceived value of the system. Three open questions guided the group discussions: a) ``How did you experience using the system'', b) ``What worked well, what did not?'', and c) ``If anything, what would you improve?''.

\begin{figure*}
  \centering
  \includegraphics[width=1.0\linewidth]{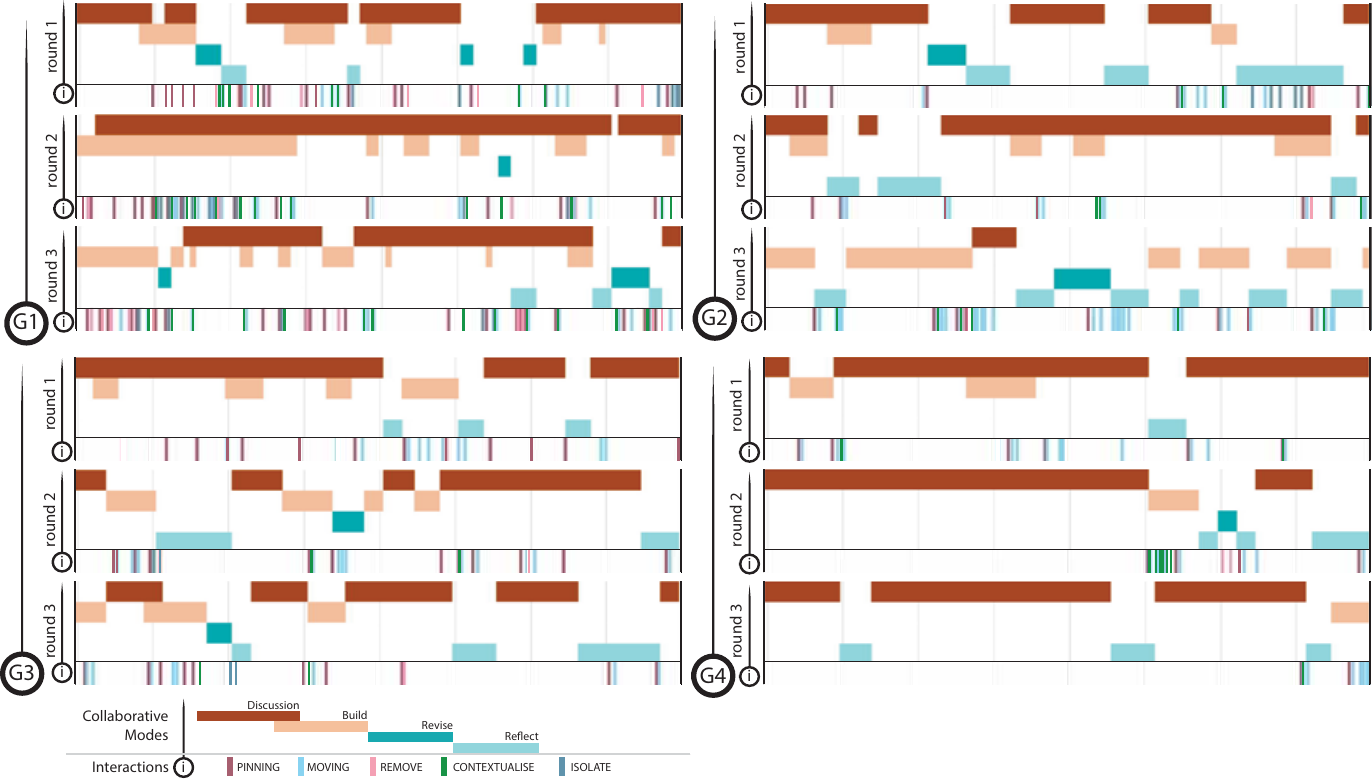}
  \caption{Collaborative Modes and interactions for all 8-minute rounds per group detailing how CollEagle supported collaborative discussion by enabling groups to shift between active discussions, building and revising their shared overview, and reflecting on this while remaining engaged with one another.}
  \Description{This figure shows in great detail, for all 8-minute rounds per group, logged interactions and observed collaborative modes, with G1 having the most logged interactions and G4 the least, while both engaged the most in discussion modes.}
    \label{fig:graph_modes}
\end{figure*}

\subsection{Data Collection}
Each study session and subsequent group discussions were recorded with the consent of the participants. Audio and video recordings, transcriptions, extracted key phrases, and screen captures taken after each interaction were collected through the CollEagle system. Quantitative data on the number of interactions and curated key phrases for each round were manually logged using screen captures and recorded video material. Failed interactions, due to two participants interacting simultaneously, and those following to undo its effects, were omitted from the final interaction log. We used screen captures to construct an overview that details the distribution of tangible annotations at the end of each round. Audio recordings for all rounds and following semi-structured group discussions were manually transcribed to gather insights from the participants' perspectives and evaluate research observations.

\subsection{Analysis} 
The video recordings were analysed and encoded into four collaborative modes: discuss, build, revise, and reflect. Our encoding is derived from collaborative sensemaking phases - extract, cluster, record, connect, review - as described by Vogt et al ~\cite{vogt2011co}, processes of gaining insight - 'provide overview', 'adjust', 'detect patterns', 'match mental model' - defined by Yi et al.~\cite{Yi2008understandinginsights}, collaborative coupling styles ~\cite{tang2006} and collaboration styles~\cite{Brudy2018overview} and construed to detail the most discernable activities in using CollEagle. The initial encoding of the video data was validated against the interaction log and revised on the basis of congruency with interaction types. We further evaluated the revised chart based on consecutive interactions of different participants to ensure that the modes comprised joint activities. The resulting graphs detail four modes to outline the collaboration of the participants while using CollEagle:

\begin{itemize}
\item[{\bfseries A}]  {\bfseries : Discuss - } comprises \textit{active} discussions
\item[{\bfseries B}]  {\bfseries : Build - } comprises activities in which \textit{all participants} collaboratively build their overview while this was the subject of discussion, or activities in which an overview was built by more than one participant during active discussions
\item[{\bfseries C}]  {\bfseries : Revise - } comprises activities in which \textit{all participants} reflected on the overview of pinned subjects and made changes while this was the subject of discussion.
\item[{\bfseries D}]  {\bfseries : Reflect - } comprises activities in which \textit{all participants} made use of the interface to reflect on their progress, either supported by subjects pinned on the interface or virtual post-its passing by, while this was the subject of discussion.  \newline
\end{itemize}

\noindent Finally, we plotted the automatically produced materials by performing network analysis~\cite{hevey2018network}. Nodes in the constructed graphs represent material CollEagle extracted from participants' conversations, with links detailing materials extracted from the same transcribed segments. With these providing insight into the distribution of topics per session for each group and across all rounds, the graphs were reconstructed to exclude network components which had 6 or fewer subjects, focussing on relationships between materials in larger components. We highlighted nodes within these graphs that comprised content curated by participants to gain insight into the relationship between externalisations produced vs. used (see Figure \ref{fig:graph_curated}).

\subsection{Observations}
\begin{figure*}
  \centering
  \includegraphics[width=1.0\linewidth]{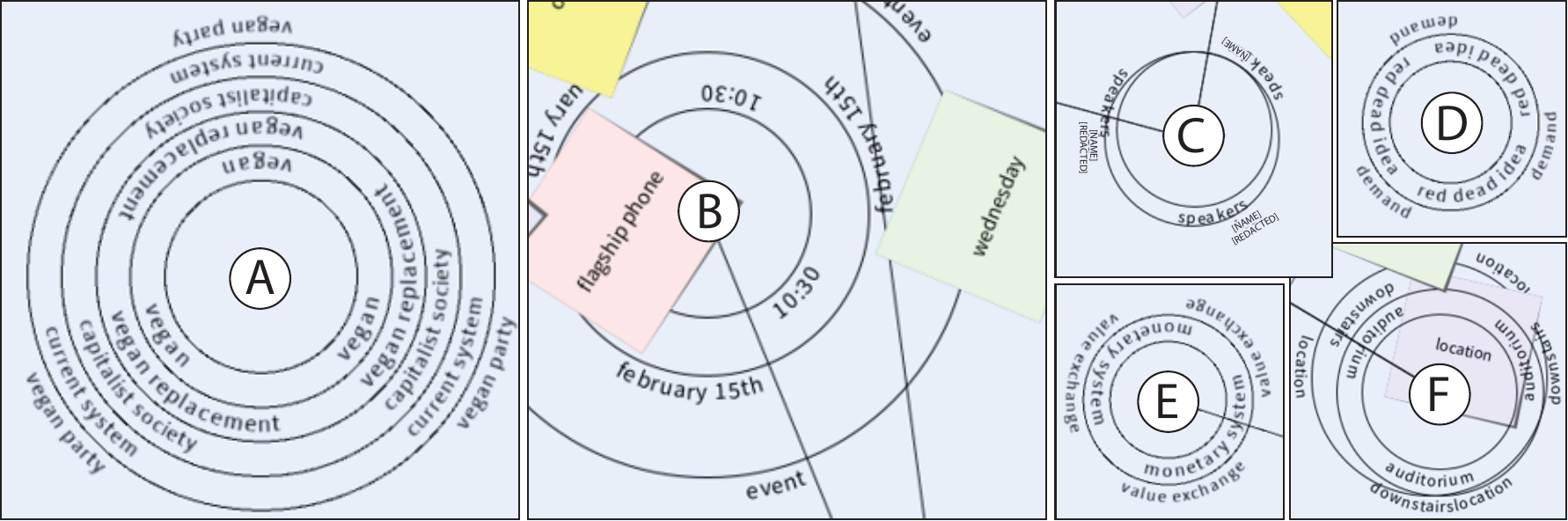}
  \caption{Virtual components to ``contextualised'' tangible annotations resembling a (A) political ``vegan party'' [G4, r2], (B) Specified deadline [G1, r3], (C) Person in charge of task [G1, r2], (F) Event location [G1, r2]. In E (inner), content was inaccurately transcribed, yet curated, resulting in refined annotation (D) for identified problem (D, inner), consequently resulting in interpreted context to this (E, outer).}
  \Description{This figure shows virtual components to ``contextualised'' tangible annotations resembling a (A) highly contextualised tangible annotation representing a political ``vegan party'' [G4, r2], (B) a specified deadline [G1, r3], (C) Person in charge of task [G1, r2], (F) Event location [G1, r2]. Lastly, a tangible annotation with inaccurately transcribed, yet curated, content (E, inner), resulting in refined annotation (D) for identified problem (D, inner), consequently resulting in interpreted context to this (E, outer).}
    \label{fig:tanganot}
\end{figure*}

Our observations provide insight into how CollEagle, enabling users to collaboratively construct shared representations using automatically produced materials, supports collaboration. Synthesising insights, reporting on observations and participants' experiences, we highlight the mediating effect of the shared interaction approach interaction enabled by the tangible interactive tabletop system. We included instances of errors and accuracies to provide insight into how participants leveraged these to support collaboration. We annotated user responses retrieved from the semi-structured group discussions by including their group and participant number (e.g., ``P3 (G1)''), for which responses elicited in using CollEagle during collaboration rounds are accompanied with the corresponding round number (e.g., ``P12 (G3), round 3''). We included figures that detail collaborative modes and interactions in each round (Figure \ref{fig:graph_modes}), participant interaction frequency per group, the ratio between transcript word count, produced material, and used material, sections of network graphs with highlighted notes of tangible annotations, and include figures of intriguing instances that highlight the purposeful use of tangible annotations (Figure \ref{fig:tanganot}). 

\subsubsection{Shared Resources}
The continuous flow of virtual post-its influenced collaboration considerably due to these reflecting the ongoing discussions. This supported participants to return to what they recently discussed and explore this as a resource for externalisation: \textit{``What else, what do we have?'' - P11 (G3, round 3)}. Presenting materials sourced from their discussion, the stream was perceived as a shared resource that captured combined efforts:  \textit{``It's like a collective thought stream or idea stream.'' - P9 (G2)}. With recently discussed topics appearing in the stream, the participants reflected on tangible annotations in the context of these: \textit{``Education, addiction, what fits into this?'' - P8 (G2, round 2)}. Consequentially, both refined versions and duplicates of materials appeared in the stream, for which some participants feared the latter could lead to cyclical discussions: \textit{``You shouldn't read out loud because then you get stuck in a loop.'' - P8 (G2, round 2)}. 

\subsubsection{Automated Annotation}
Participants described the collocated collaboration as both dynamic and more efficient, particularly because the system negated the need for manual note-taking: \textit{``During brainstorming, it was quite nice because normally you just put Post-its down, and now it already does it for you.'' - P8 (G2)} -- \textit{``It feels more dynamic, and it feels quicker because you're not actively focusing on what you're writing down.'' - P10 (G3)}. This annotative support was widely appreciated by participants, with these imagining instances where the system's annotating capabilities would have been advantageous, drawing from personal preferences, everyday experiences, and professional backgrounds: \textit{``It's like dynamic minutes.'' - P3 (G1)} -- \textit{``In meetings with my supervisors, we are constantly talking about a lot of ideas, and they will mention a lot of names and works. Using this, taking notes can be just a simple interaction, then I can really record it'' - P15 (G4)}. 

\subsubsection{Mediating Stream}
The presence of the stream of virtual post-its influenced how participants interacted with each other. Reflecting on how they made use of the stream of collaborative materials, participants touched on strategies with some describing distinct approaches: \textit{``Sometimes I'm just scanning, and other times I'm looking for something I want to get stuck.'' - P13 (G3)}. With the influx of post-its actuated by ongoing discussion, we observed the discussion of Group 3 subsided in collectively focusing on the interface to inspect virtual post-its. As this led to the gradual disappearance of materials, the absence of these prompted Group 3 to continue their discussions: \textit{``Let's talk more, instead of trying...'' - P12 (G3, round 1)}.   

\subsubsection{Temporality and Uncertainty}
The temporal availability of materials presented in the stream created a sense of immediacy. Feeling uncertain about whether the content would reappear stimulated  some participants to interact: \textit{“There is a time factor, like, I don’t know if this word is going to come up. So if I think this is interesting, I have to take it, I have to pick it.“ - P8 (G2)}. At the same time, this uncertainty stimulated participants to engage in grounding, with these repeating recent remarks to ensure a post-it would appear: \textit{“Oh nice word, ’bio-receptive”’- P11 (G3, round 2)}.  Correspondingly, participants often held on to artefacts, readying them for placement on the interface or seeking peer assistance to pin items out of reach on the interface. 

\subsubsection{Externalisation Techniques}
Interactions to collect additional post-its using existing tangible annotations, `contextualising' the previously attached information, occurred in close proximity to one another (see Figure \ref{fig:graph_modes}). Beyond instances where these comprised interactions resulting in refined annotations, we observed participants experimenting with strategies, collecting numerous annotations using a single artefact to encapsulate complex ideas (see Figure \ref{fig:tanganot}:A). Taking this a step further, we observed Group 1 using tangible annotations purposefully employing tangible representations to represent team members and ``contextualising'' these with corresponding tasks and locations (see Figure \ref{fig:tanganot}:C~\& \ref{fig:tanganot}:F, respectively). Relating to this, we observed G3 referring to tangible annotations as their ``pillars''.

\subsubsection{Interaction Techniques}
Combined with the use of physical artefacts, the system's ability to automatically connect related subjects, further shaped and influenced collaboration, with P10 ``isolating'' one of these, eliciting discourse about how these should be interpreted by the group (see Figure \ref{fig:graph_modes}:G3-r2): \textit{"Yeah, but let's see it as connected to everything, not something separate - P11 (G3, round 2)"}. Because manually drawing lines between tangible annotations was not supported, G3 resorted to clustering these: \textit{``I want to connect this'' - P11 -- ``Just put them close to each other'' - P10 (G3, round 1)}. Taking an exploratory approach to resolve this, participants in Group 1 intuitively stacked one tangible annotation on top of another, often using connections as suggestions to do so. This interaction was not taken into account and resulted in misaligned digital components (see Figure \ref{fig:tanganot}:C and \ref{fig:tanganot}:F), with connections of the artefact on top remaining intact. Surprisingly, moving the stack to another place caused the digital components to align with an unusually resized diameter.

\subsubsection{Interacting towards consensus}
Situated in the direct periphery of participants, we observed how interacting with the system and making tangible annotations became a way to confirm or support statements: \textit{``I like what you said, that's why I tried to catch it a bit'' - P7 (G2, round 1)}. Here, the stark contrast between the number of interactions and the tangible annotations made by G4, compared to the other groups (see Figure \ref{fig:graph_modes}), suggests the frequency with which participants interact with the system signals an extent to which participants can agree, and perhaps captures the pace of their collaborative effort. That is, unable to reach a consensus on central issues in the complex societal problem, G4 spent most of their time discussing (see Figure \ref{fig:graph_modes}) continuing their initial task in later rounds: \textit{``Let's just say soy milk is better, just for the sake of moving forward'' - P16 (G4, round 3)}. 

\subsubsection{Roles and Responsibilities}
\begin{figure}[hbp]
  \centering
  \includegraphics[width=1.0\linewidth]{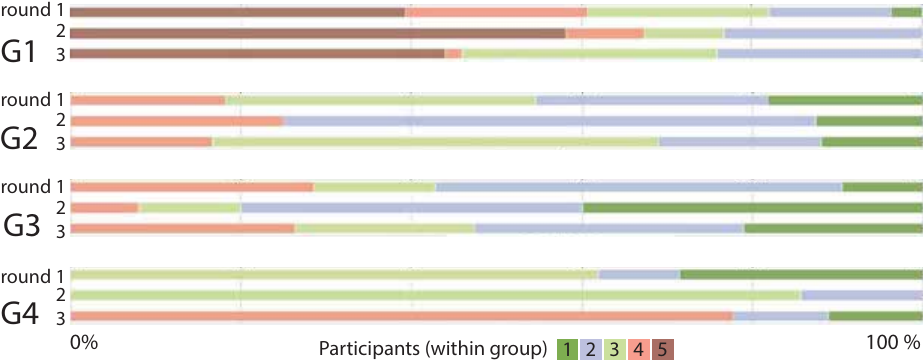}
  \caption{Distribution of interactions per participant, expressed as each participant’s share of the total interactions within their group for the corresponding round.}
  \Description{This figure displays the percentage of interactions with CollEagle per participant, calculated as the total number of interactions per participant divided by the total number of interactions from all the group participants for the corresponding round. Aside from a few rounds, most appear to indicate participation through interaction was balanced.}
    \label{fig:participationpercentage}
\end{figure}

The collaborative practice enables by CollEagle mediated roles and responsibilities among participants, with these shifting between discussing and interacting. \textit{``It's kind of like, someone is talking, and then one person or two people are just focused here and trying to pinpoint everything.'' - P3 (G1)}. Participants compared this with common approaches to collaborative activities, highlighting the shared nature of collaborative practices emerging from the use of the system. \textit{``In a traditional setting with whiteboards, sticky notes, etc, you kind of switch between talking and writing - and it always kind of feels like you have a designated leader to be the writer.'' - P11 (G3)}. With externalised materials resulting from shared efforts, engaging with the system was perceived as a way to gain insight into roles within the group: \textit{``It sort of emphasises team dynamics or team roles. I can imagine you can start to see really quickly who is making the decisions.`` - P8 (G2)}. Supporting this claim, Figure \ref{fig:participationpercentage} shows the secretary (P3) and chairman (P5) of Group 1 interacted more often than their peers. Notably, while interaction frequency for Group 1 appears balanced for the first round, P3 and P5 were responsible for 60\% of interactions, with this increasing in using CollEagle to support their scheduled board meeting in rounds that followed. 

\subsubsection{Collaboration Strategies} 
Bridging their output across rounds, we observed that groups briefly summarised and discussed the round before, verbalising the main topics to have them appear on the interface (see Figure \ref{fig:graph_modes}), with tangible annotations constructed from these recaps forming a foundation to continue their collaborative effort: \textit{``Well, we have our different pillars now, we have service, products, material and politics'' - P10 (G3, round 2)}. In contrast to the efficacy of verbalising topics conversationally for this to appear on the interface, the participants of G1, aiming to construct a detailed plan specifying times and dates (see Figure \ref{fig:tanganot}:F), explicitly stated these repeatedly. Seeing Group 1 even shouting specified dates and waiting for these to show, resulting in an abundance of post-its with similar contents, suggests the interactivity provided by the system is specifically beneficial in supporting collaboration when efforts are guided by discussion.

\begin{figure}
  \centering
  \includegraphics[width=1.0\linewidth]{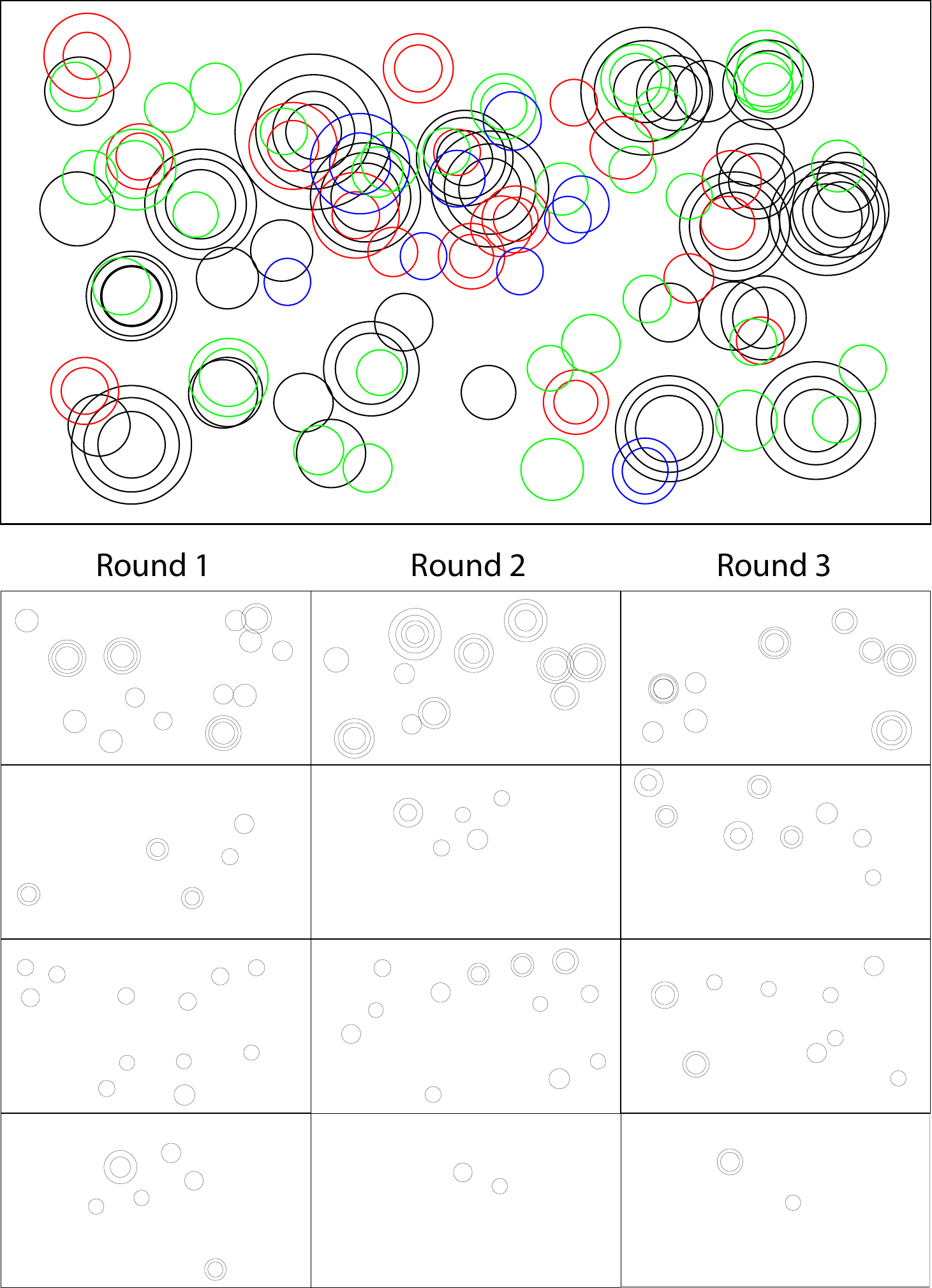}
  \caption{Combined distribution of tangible annotations on the interface after all group rounds, with G1=black, G2=red, G3=green, G4=blue (top). Distribution of tangible annotations after each round per group, starting with G1 and ending with G4 (bottom). Multiple bounds comprise contextualised annotations}
  \Description{This figure shows the distribution of tangible annotations on the interface at the end of all group rounds.}
    \label{fig:dist}
\end{figure}

\subsubsection{Perceived Workspace}
Interacting with the system rarely occurred in single instances, with `pinning' interactions often rapidly followed by `move' or `remove' interactions (see Figure \ref{fig:graph_modes}). Initially, due to the circular movement of the post-its, and the amount of these populating the display, empty spaces to construct overviews appeared to be scarce: \textit{“Like we have empty space here, here and here.” – P1 (G1)}. Notably, the perceived workspace gradually expanded, with the areas occupied by the stream paths deemed viable once `empty spaces' were populated (see Figure 11). Reflected in how tangible annotations were augmented in the stream (see Figure \ref{fig:tanganot}: B, C and D), the dynamic nature of the interface allows for this expansion. That is, with post-its moving along their paths and text rotating around the artefact, participants were able to revisit tangible annotations in the paths with ease, despite these temporarily overlapping with post-its (see Figure \ref{fig:tanganot}:B:F).

\begin{figure*}
   \centering
   \includegraphics[width=1.0\linewidth]{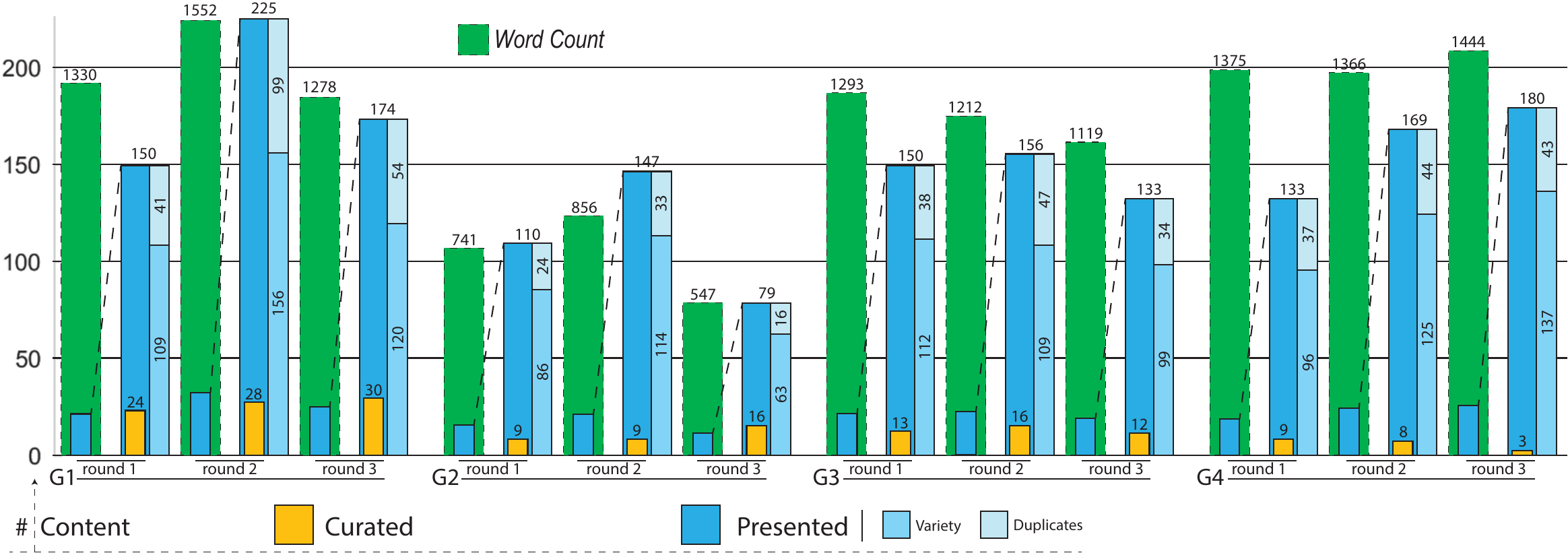}
   \caption{Quantitative data for each round, showing the total number of words transcribed (green), amount of content (i.e. post-its) presented (blue), including the variety of content presented and number of duplicates, and amount of content curated (orange).}
   \Description{This figure shows quantitative data for each round, showing the total number of words transcribed (green), amount of content (i.e. post-its) presented, the variety of content presented and number of duplicates (blue), amount of content curated (orange).}
     \label{fig:quantmech}
 \end{figure*}

 \begin{figure*}
  \centering
  \includegraphics[width=\linewidth]{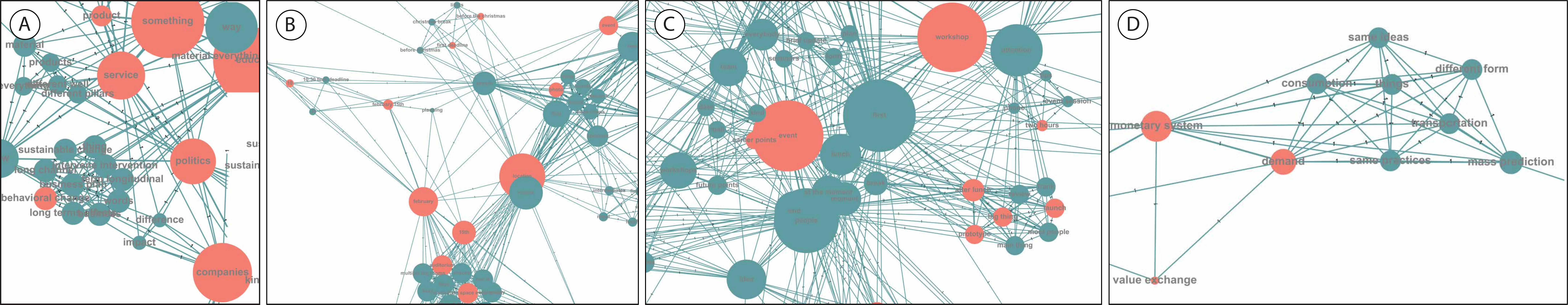}
  \caption{Sections of constructed network graph showing curated content (red) positioned within dense networks of topics (A,B,C), suggesting content is curated to highlight central concepts or add nuance and provide contexts to these (B,C,D) in support of a groups’ shared understanding.}
  \Description{This figure shows sections of constructed network graph to demonstrate how CollEagle's interactivity holds the potential to enable novel approaches in collaboration analytics, showing showing curated content (red) positioned within dense networks of topics}
    \label{fig:zoomed}
\end{figure*}

\begin{figure*}
\centering
  \includegraphics[width=0.8\textwidth, trim= 35 35 35 35, clip]{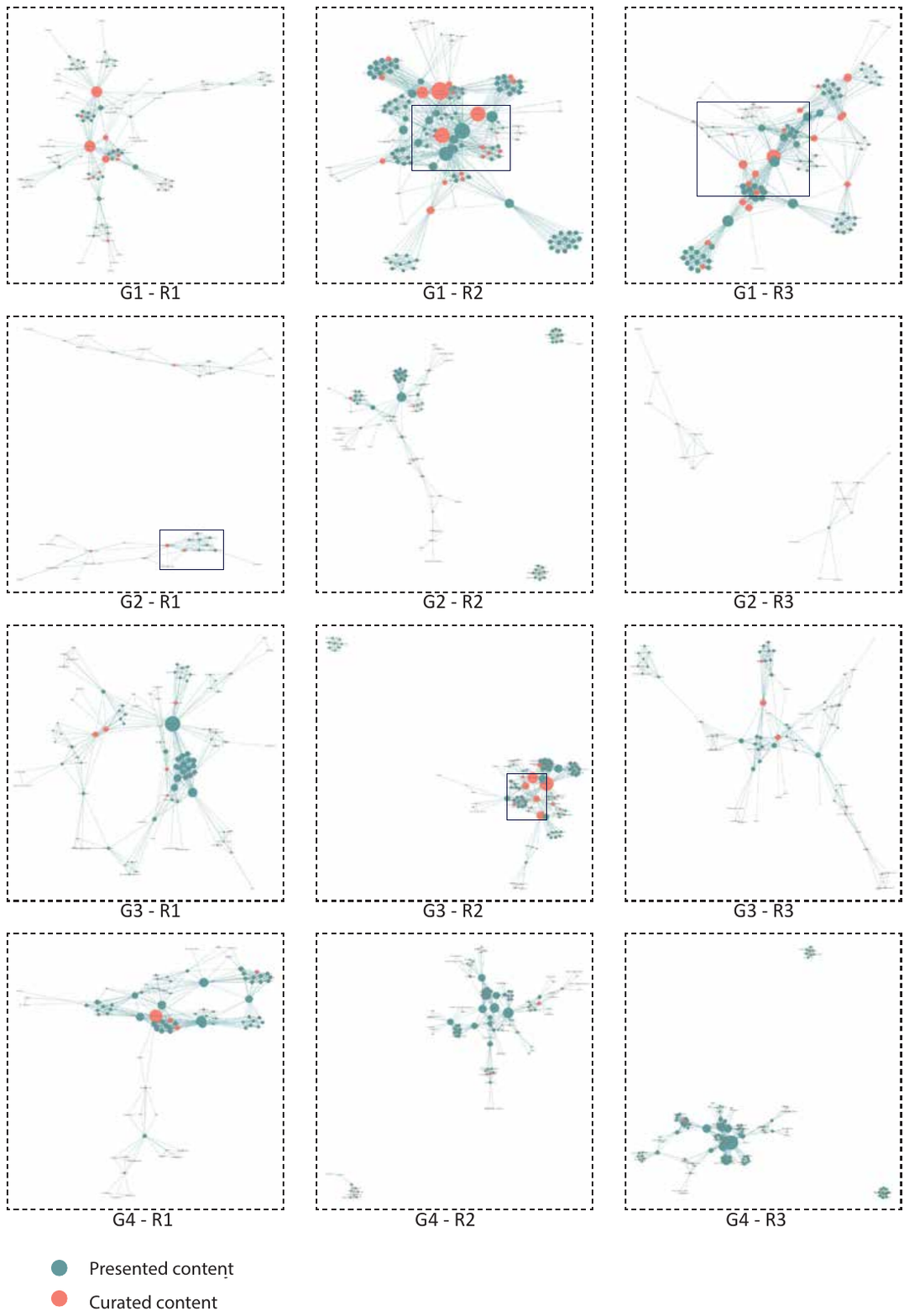}
  \caption{Network analysis of extracted content per round for each group, detailing origins for Figure \ref{fig:zoomed}.}
  \Description{This figure shows the results of the network analysis for the subjects extracted per round for each group, detailing zoomed sections presented in Figure \ref{fig:zoomed}.}
    \label{fig:graph_curated}
\end{figure*}

\subsubsection{Diverging through Inaccuracy}
With participants being non-native speakers, inaccurately transcribed content was inevitable. The post-its produced as a result of inaccuracies were mostly disregarded or had a comedic value. However, at times, these caused the topic of discussion to diverge in making sense of these. The most pertinent effect was observed in Group 2, with P9 selecting inaccurately transcribed material, thereby emphasising their intent to discuss the topic they derived from this:  \textit{"This says 'red dead idea', is that something someone said? Does this refer to Marxist economies?" - P9 (G2, round 1)}. Consequently, both the tangible annotations that initially framed their problem space and the one selected by P9 were refined, with one detailing a more distinct framing and the other representing a highly abstracted concept (see Figures \ref{fig:tanganot}:E:D and \ref{fig:zoomed}:D).

\subsubsection{Implementation for Turn-taking}
As participants frequently interacted simultaneously, the digital components of tangible annotations remained on the interface, by which our implementation for turn-based interaction impacted the quality of collaboration. Most groups quickly learnt that returning artefacts to their previous location allowed for performing the interactions sequentially. In contrast, emphasising how intentionality in interacting was mediated by physically situated materials, Group 3 decided persisting visualisations paired with physical components now carried more weight \textit{``Alright, they'll just stay there, they're not relevant right now because they do not have a block on it.'' (P10)  -- ``Yeah, exactly'' (P13) }.



\subsubsection{Highlighting in Curation}
Figure~\ref{fig:graph_curated} shows the network analysis of the content extracted per round, with the curated content highlighted in red. In the case of G1, the constructed graph suggests the group curated content that continued to be central to their collaborative discussions (see Figure~\ref{fig:graph_curated}). A similar structure in the graph and highlights can be seen for G3, referring to tangible annotations as their ``pillars'' during their second round (see Figure \ref{fig:zoomed}:A). In most other cases, we see that the curated topics are distributed across somewhat smaller clusters, away from the central, most verbalised topics. In these cases, G1 curated content to capture decisions and document outcomes (see Figures \ref{fig:tanganot}:B:C:F \& \ref{fig:zoomed}:C)), while G2 and G3 curated content to provide context and add nuance (see Figure \ref{fig:zoomed}: D) to externalise topics of discussion and maintain a shared understanding of these (see Figure \ref{fig:tanganot}:D).

\subsubsection{Ratio of Produced, Duplicate and Used Materials}
Most of the presented content consisted of a post-it containing a single word, 28\% consisting of two, and a few of three or more. Participants interacted with a small percentage of the produced materials (see Figure ~\ref{fig:quantmech}), and selected these with a ratio ranging from 0.02 (G4, round 3) to 0.20 (G2, round 3). Discussion and reviewing the content presented in the stream caused duplicates to appear (median ratio of materials presented once = 0.65). This effect appeared to be most prevalent in Group 1 (median ratio of duplicate materials = 0.44), followed by Group 4 (median = 0.35), Group 3 (median = 0.34) and Group 2 (median = 0.28).



\section{Discussion}
This work proposes a \textit{semantic interaction mechanism}, comprising two interactive loops, that specifies how an implicitly created content stream can be mediated by explicit interactions to support collaborative activities. We foreground conversations and debate as the crux of collocated collaboration and suggest that designing collocated systems starts with supporting humans interacting. Introducing semantic interactivity, the design and focus of CollEagle shifts the core interaction approach from \textit{creating individual work} and sharing this with others to a \textit{collective data stream} that enables curating and configuring shared materials through collaborative actions.

\subsection{From Creation to Curation} 
Our study suggests that `curation of content' might be a more suitable interaction approach for ambiguous and complex forms of conversation-centric collocated collaboration than the classic 'create and share content'. While in CollEagle we used a tabletop setup for this form of curation in collaboration, we propose that the shared interaction approach translates to other device form-factors to enable collaborative content creation, such as wall-mounted touch displays~\cite{mandryk2002display}, hybrid setups~\cite{houben2014activityspace,Brudy2018overview} and augmented reality systems~\cite{Wells2020CollabAR} -- or even constellations of mobile devices~\cite{Brudy2019CrossDevice}.

\subsubsection{Collective Data Stream} 
In the collaboration support enabled by semantic interactivity, data is produced implicitly~\cite{schmidt2000implicit} through conversation, and the main interaction approach is to curate and collect relevant information~\cite{brudy2016curationspace}. This reduces the `cold start' problem when interacting with shared surfaces during collaboration, as content and data flow as soon as the collaborative discussion commences and proceeds to match the flow of conversation as it continues. Because there is no need to conduct configuration or preparation work to translate parts of the collaboration and conversation into a digital representation, semantic interactivity greatly reduces the workload, configuration work, and general overhead of using collocated devices ~\cite{houben2014activityspace}. Further, it makes known problems with shared surfaces~\cite{wallace2017disappearing}, such as the need to support transitions from individual activities to shared content~\cite{scott2003system}, redundant. That is, we observed that this approach caused fewer distractions or context switches than what was reported in prior work, where the role of devices in collocated collaboration typically led to context switches or interruptions~\cite{scott2003system,Wells2020CollabAR,homaeian2022handoff,Brudy2018overview}. Specifically, shared focus on the group task remained intact as any distracting feats had individuals focus attention on externalisations that directly reflected the ongoing conversation.

\subsubsection{Grounding} 
Our findings show that semantic interactivity supports a shared approach to content creation, creating a medium for users to remain engaged with each other \textit{through} interacting with this content. Explicit writing tools and input devices were replaced by implicit context-aware voice input, reducing configuration work and shared overhead, and focussing shared attention on person-to-person collaboration. Highlighting its ability to leverage computer-mediated productivity tools in ways that foster the unique qualities of collocated collaboration, our collaboration method enabled a new form of grounding~\cite{clark1991grounding}, where engaging with the system unintentionally affirmed what was understood by the group. These \textit{grounding effects} emphasise how the inherent task of participating in a discussion is to continually and repeatedly reach a point of shared understanding. 



\subsubsection{Situated sense-making} 
Embedded in a tangible interactive tabletop system, the explicit interaction loop actuated by the use of tangibles stimulated \textit{collaborative action} on a \textit{shared interface}. A participatory sensemaking practice emerged as assigning meaning to artefacts supported enhanced forms of sensemaking, confirming insights from previous work~\cite{jaasma2017}. Expanding on this, with the meaning ascribed to tangibles persisting through its augmented component, tangible annotations enabled artefacts to represent increasingly complex matters and became situated externalisations. The resulting tangible annotations then acted as spatially stable semantic anchors~\cite{houben2014activityspace}, embodying concepts that resembled the shared understanding of users, for which interacting with this further added context to their meaning. Although initially causing unsuccessful interactions, the turn-based interaction format of the system eventually increased participants' awareness of each other's actions, confirming reports of previous work~\cite{morris2006}. This increased awareness translated into more deliberate actions and discussions about what content to pin on the interface, exemplifying the interface's role in enhancing attention to collaborative action~\cite{Messeguer2008,smit2022sense}. 

\subsubsection{Decision-making} 
In addition to signalling agreements, grounding, and sensemaking, interacting with the interface was perceived as a form of \textit{decision-making}. This is due to the positioning and omnidirectional nature of the interface, considering how previous works~\cite{kruger2004roles} describe that for tabletop interfaces, interactions to rotate content and territoriality~\cite{scott2004territoriality} play a role in supporting collaboration. That is, the omnidirectional interface is designed to enable 'democratic' viewing from all sides, thus omitting the need for rotating or assigning content to physical directions or private spaces. Moreover, because the content persisting on the interface directly results from discussions, combined with the size of the display, this also seems to counteract territoriality on the level of individuals and emphasises \textit{shared ownership of the interface}. These effects are further emphasised through the implicit interaction~\cite{serim2019} approach for creating content, omitting territoriality perceived through the use of virtual keyboards or additional devices as reported in prior work~\cite{Klinkhammer2018tabletop}, suggesting content appearing on the interface was truly \textit{perceived to result from a shared effort.} \newline 

\noindent In summary, leveraging NLP to enable interactive annotation streams in collaborative systems and settings can profoundly reshape interactions, discussions, and outcomes. Our implementation limits this work to clustering and combining annotations to support the creation of shared representations, for which CollEagle's role as a mediator shows the potential for semantic interactivity to enable the integration of new and existing tools to enhance, challenge, and ultimately transform the collaborative landscape.

\section{Limitations}
This paper aims primarily to enable shared interfaces through group-level interactivity for collocated collaboration -- integrating the efficiency of digital tools with the communicative benefits of face-to-face collaboration. Although the presented system supports tangible interactions, analysing their effects was not the central aim; insights in this area were limited, and including them would detract from the main contribution. Similarly, although natural language processing (NLP) was implemented to support a shared interaction model, the work does not report performance metrics such as Word-Error-Rate. We argue that collaborative systems should anticipate scenarios where users lack language fluency or have speech impediments. Therefore, our focus is on understanding the implications of incorrectly transcribed content and the broader consequences of using NLP regardless of model accuracy.

\section{Conclusion}
This paper introduces \textit{semantic interactivity}, a shared interaction approach that emerged through the design and development of CollEagle. This interactive tabletop system enables a shared externalisation practice by integrating \textit{NLP} with \textit{tangible interactions}. CollEagle merges \textit{computational conveniences} with \textit{the quality of sharing presence} with two interaction loops, situated in the \textit{digital} and \textit{physical}. A \textit{shared interaction mechanism} surfaced that supports joint activities by using natural language processing (NLP) to continuously generate snippets of text, while providing a low-effort method to create, organise, and structure these, for users to capture and support ongoing discussions. The preliminary study demonstrates that CollEagle effectively mediates collaboration by enabling a dynamic annotation method -- highlighting the potential for semantic interactivity to enable novel interaction techniques to support joint activities.



\bibliographystyle{ACM-Reference-Format}
\bibliography{main}

@article{al2018review,
  title={A review of brainstorming techniques in higher education},
  author={Al-Samarraie, Hosam and Hurmuzan, Shuhaila},
  journal={Thinking Skills and creativity},
  volume={27},
  pages={78--91},
  year={2018},
  publisher={Elsevier}
}

@article{barki2001small,
  title={Small group brainstorming and idea quality: Is electronic brainstorming the most effective approach?},
  author={Barki, Henri and Pinsonneault, Alain},
  journal={Small Group Research},
  volume={32},
  number={2},
  pages={158--205},
  year={2001},
  publisher={Sage Publications Sage CA: Thousand Oaks, CA}
}

@incollection{adan2023a,
  title = {{{CollEagle}}; {{Tangible Human-AI Interaction}} for {{Collocated Collaboration}}},
  booktitle = {{{HHAI}} 2023: {{Augmenting Human Intellect}}},
  author = {Adan, Olaf and Houben, Steven},
  year = {2023},
  pages = {416--418},
  publisher = {IOS Press},
  address = {Amsterdam, The Netherlands},
  doi = {10.3233/FAIA230115},
  langid = {english}
}

@article{horvitz1999uncertainty,
  title={Uncertainty, action, and interaction: In pursuit of mixed-initiative computing},
  author={Horvitz, Eric},
  journal={IEEE Intelligent Systems},
  volume={14},
  number={5},
  pages={17--20},
  year={1999}
}

@article{allen1999mixed,
  title={Mixed-initiative interaction},
  author={Allen, James E and Guinn, Curry I and Horvtz, Eric},
  journal={IEEE Intelligent Systems and their Applications},
  volume={14},
  number={5},
  pages={14--23},
  year={1999},
  publisher={IEEE}
}

@article{bardram2005activity,
  title={Activity-based computing: support for mobility and collaboration in ubiquitous computing},
  author={Bardram, Jakob E},
  journal={Personal and Ubiquitous Computing},
  volume={9},
  pages={312--322},
  year={2005},
  publisher={Springer}
}

@inproceedings{scott2003system,
  title={System guidelines for co-located, collaborative work on a tabletop display},
  author={Scott, Stacey D and Grant, Karen D and Mandryk, Regan L},
  booktitle={ECSCW 2003: Proceedings of the Eighth European Conference on Computer Supported Cooperative Work 14--18 September 2003, Helsinki, Finland},
  pages={159--178},
  year={2003},
  organization={Springer}
}

@article{schmidt2000implicit,
  title={Implicit human computer interaction through context},
  author={Schmidt, Albrecht},
  journal={Personal technologies},
  volume={4},
  pages={191--199},
  year={2000},
  publisher={Springer}
}

@inproceedings{Bergstrom2009,
author = {Bergstrom, Tony and Karahalios, Karrie},
title = {Conversation clusters: grouping conversation topics through human-computer dialog},
year = {2009},
isbn = {9781605582467},
publisher = {Association for Computing Machinery},
address = {New York, NY, USA},
url = {https://doi.org/10.1145/1518701.1519060},
doi = {10.1145/1518701.1519060},
booktitle = {Proceedings of the SIGCHI Conference on Human Factors in Computing Systems},
pages = {2349–2352},
numpages = {4},
location = {Boston, MA, USA},
series = {CHI '09}
}

@inproceedings{scott2004territoriality,
  title={Territoriality in collaborative tabletop workspaces},
  author={Scott, Stacey D and Carpendale, M Sheelagh T and Inkpen, Kori},
  booktitle={Proceedings of the 2004 ACM conference on Computer supported cooperative work},
  pages={294--303},
  year={2004}
}

@inproceedings{Chen2023,
author = {Chen, Yingting and Kanno, Taro and Furuta, Kazuo},
title = {Cognition-Oriented Facilitation and Guidelines for Collaborative Problem-Solving Online and Face-to-Face: An in-Depth Examination of Format and Facilitation Influence on Problem-Solving Performance},
year = {2023},
isbn = {9781450394215},
publisher = {Association for Computing Machinery},
address = {New York, NY, USA},
url = {https://doi.org/10.1145/3544548.3581112},
doi = {10.1145/3544548.3581112},
booktitle = {Proceedings of the 2023 CHI Conference on Human Factors in Computing Systems},
articleno = {61},
numpages = {15},
location = {Hamburg, Germany},
series = {CHI '23}
}

@inproceedings{jung2017,
  title = {Possibilities and {{Limitations}} of {{Online Document Tools}} for {{Design Collaboration}}: {{The Case}} of {{Google Docs}}},
  shorttitle = {Possibilities and {{Limitations}} of {{Online Document Tools}} for {{Design Collaboration}}},
  booktitle = {Proceedings of the 2017 {{ACM Conference}} on {{Computer Supported Cooperative Work}} and {{Social Computing}}},
  author = {Jung, Young-Wook and Lim, Youn-kyung and Kim, Myung-suk},
  year = {2017},
  month = feb,
  series = {{{CSCW}} '17},
  pages = {1096--1108},
  publisher = {Association for Computing Machinery},
  address = {New York, NY, USA},
  doi = {10.1145/2998181.2998297},
  urldate = {2025-05-27},
  isbn = {978-1-4503-4335-0}
}

@inproceedings{Jensen2018digitizingtools,
author = {Jensen, Mads M\o{}ller and R\"{a}dle, Roman and Klokmose, Clemens N. and Bodker, Susanne},
title = {Remediating a Design Tool: Implications of Digitizing Sticky Notes},
year = {2018},
isbn = {9781450356206},
publisher = {Association for Computing Machinery},
address = {New York, NY, USA},
url = {https://doi.org/10.1145/3173574.3173798},
doi = {10.1145/3173574.3173798},
booktitle = {Proceedings of the 2018 CHI Conference on Human Factors in Computing Systems},
pages = {1–12},
numpages = {12},
location = {<conf-loc>, <city>Montreal QC</city>, <country>Canada</country>, </conf-loc>},
series = {CHI '18}
}

@inproceedings{Houben2013,
author = {Houben, Steven and Bardram, Jakob E.},
title = {ActivityDesk: multi-device configuration work using an interactive desk},
year = {2013},
isbn = {9781450319522},
publisher = {Association for Computing Machinery},
address = {New York, NY, USA},
url = {https://doi.org/10.1145/2468356.2468484},
doi = {10.1145/2468356.2468484},
booktitle = {CHI '13 Extended Abstracts on Human Factors in Computing Systems},
pages = {721–726},
numpages = {6},
location = {Paris, France},
series = {CHI EA '13}
}

@ARTICLE{morris2006,
  author={Morris, Meredith Ringel and Cassanego, Anthony and Paepcke, Andreas and Winograd, Terry and Piper, Ann Marie and Huang, Anqi},
  journal={IEEE Computer Graphics and Applications}, 
  title={Mediating Group Dynamics through Tabletop Interface Design}, 
  year={2006},
  volume={26},
  number={5},
  pages={65-73},
  doi={10.1109/MCG.2006.114}}

@inproceedings{tang2006,
author = {Tang, Anthony and Tory, Melanie and Po, Barry and Neumann, Petra and Carpendale, Sheelagh},
title = {Collaborative Coupling over Tabletop Displays},
year = {2006},
isbn = {1595933727},
publisher = {Association for Computing Machinery},
address = {New York, NY, USA},
url = {https://doi.org/10.1145/1124772.1124950},
doi = {10.1145/1124772.1124950},
booktitle = {Proceedings of the SIGCHI Conference on Human Factors in Computing Systems},
pages = {1181–1190},
numpages = {10},
location = {Montr\'{e}al, Qu\'{e}bec, Canada},
series = {CHI '06}
}

@inproceedings{Plaue2009ConferenceRoomAsToolBox-TechAndSOcialRoutinesIncorporateMeetingSPaces,
author = {Plaue, Christopher and Stasko, John and Baloga, Mark},
title = {The Conference Room as a Toolbox: Technological and Social Routines in Corporate Meeting Spaces},
year = {2009},
isbn = {9781605587134},
publisher = {Association for Computing Machinery},
address = {New York, NY, USA},
url = {https://doi.org/10.1145/1556460.1556476},
doi = {10.1145/1556460.1556476},
booktitle = {Proceedings of the Fourth International Conference on Communities and Technologies},
pages = {95–104},
numpages = {10},
location = {University Park, PA, USA},
series = {C\&T '09}
}

@incollection{clark1991grounding,
  title = {Grounding in Communication},
  booktitle = {Perspectives on Socially Shared Cognition},
  author = {Clark, Herbert H. and Brennan, Susan E.},
  year = {1991},
  pages = {127--149},
  publisher = {American Psychological Association},
  address = {Washington, DC, US},
  doi = {10.1037/10096-006},
  isbn = {978-1-55798-121-9}
}

@inproceedings{wallace2017disappearing,
  title={The Disappearing Tabletop: Social and Technical Challenges for Cross-Surface Collaboration},
  author={Wallace, James R and Houben, Steven and Anslow, Craig and Lucero, Andr{\'e}s and Rogers, Yvonne and Scott, Stacey D},
  booktitle={Proceedings of the 2017 ACM International Conference on Interactive Surfaces and Spaces},
  pages={482--487},
  year={2017}
}

@article{homaeian2022handoff,
  title={Handoff and Deposit: Designing Temporal Coordination in Cross-Device Transfer Techniques for Mixed-Focus Collaboration},
  author={Homaeian, Leila and Wallace, James R and Scott, Stacey D},
  journal={Proceedings of the ACM on Human-Computer Interaction},
  volume={6},
  number={CSCW2},
  pages={1--23},
  year={2022},
  publisher={ACM New York, NY, USA}
}

@book{suchman1987plans,
  title={Plans and situated actions: The problem of human-machine communication},
  author={Suchman, Lucy A},
  year={1987},
  publisher={Cambridge university press}
}

@article{mandryk2002display,
  title={Display factors influencing co-located collaboration},
  author={Mandryk, Regan L and Scott, Stacey D and Inkpen, Kori M},
  journal={Comference Supplement to ACM CSCW},
  volume={2},
  pages={10},
  year={2002},
  publisher={Citeseer}
}

@article{kawakita1991original,
  title={The original KJ method},
  author={Kawakita, Jiro},
  journal={Tokyo: Kawakita Research Institute},
  volume={5},
  pages={1991},
  year={1991}
}

@book{novak2010learning,
  title={Learning, creating, and using knowledge: Concept maps as facilitative tools in schools and corporations},
  author={Novak, Joseph D},
  year={2010},
  publisher={Routledge}
}

@article{oppl2014facilitating,
  title={Facilitating shared understanding of work situations using a tangible tabletop interface},
  author={Oppl, Stefan and Stary, Christian},
  journal={Behaviour \& Information Technology},
  volume={33},
  number={6},
  pages={619--635},
  year={2014},
  publisher={Taylor \& Francis}
}

@article{novak2006theory,
  title={The theory underlying concept maps and how to construct them},
  author={Novak, Joseph D and Ca{\~n}as, Alberto J},
  journal={Florida Institute for Human and Machine Cognition},
  volume={1},
  number={1},
  pages={1--31},
  year={2006}
}

@inproceedings{oppl2009conceptmaptabletop,
author = {Oppl, Stefan and Stary, Christian},
title = {Tabletop concept mapping},
year = {2009},
isbn = {9781605584935},
publisher = {Association for Computing Machinery},
address = {New York, NY, USA},
url = {https://doi.org/10.1145/1517664.1517721},
doi = {10.1145/1517664.1517721},
booktitle = {Proceedings of the 3rd International Conference on Tangible and Embedded Interaction},
pages = {275–282},
numpages = {8},
location = {Cambridge, United Kingdom},
series = {TEI '09}
}

@inproceedings{houben2014activityspace,
  title={Activityspace: Managing device ecologies in an activity-centric configuration space},
  author={Houben, Steven and Tell, Paolo and Bardram, Jakob E},
  booktitle={Proceedings of the Ninth ACM International Conference on Interactive Tabletops and Surfaces},
  pages={119--128},
  year={2014}
}

@article{olson2002currently,
  title={The (currently) unique advantages of collocated work},
  author={Olson, Judith S and Teasley, Stephanie and Covi, Lisa and Olson, Gary},
  journal={Distributed work},
  pages={113--135},
  year={2002},
  publisher={MIT Press, Cambridge, MA}
}

@phdthesis{Houben2016,
title = "An activity-centric approach to configuration work in distributed interaction",
author = "Steven Houben",
note = "Proefschrift.",
year = "2016",
language = "English",
isbn = "978-87-7949-329-2",
series = "ITU-DS",
publisher = "IT University of Copenhagen",
address = "Denmark",
school = "IT University of Copenhagen",
}

@article{bohus2021platformPSi,
  title={Platform for situated intelligence},
  author={Bohus, Dan and Andrist, Sean and Feniello, Ashley and Saw, Nick and Jalobeanu, Mihai and Sweeney, Patrick and Thompson, Anne Loomis and Horvitz, Eric},
  journal={arXiv preprint arXiv:2103.15975},
  year={2021}
}

@misc{AzureCogn,
  author = {{Microsoft}},
  title = {Azure Cognitive Services for Language},
  year = {2022},
  url = {https://language.cognitive.azure.com/},
  date = {},
}

@inproceedings{Scott2015localremotegroupawareness,
author = {Scott, Stacey D. and Graham, T.C. Nicholas and Wallace, James R. and Hancock, Mark and Nacenta, Miguel},
title = {"Local Remote" Collaboration: Applying Remote Group AwarenessTechniques to Co-Located Settings},
year = {2015},
isbn = {9781450329460},
publisher = {Association for Computing Machinery},
address = {New York, NY, USA},
url = {https://doi.org/10.1145/2685553.2685564},
doi = {10.1145/2685553.2685564},
booktitle = {Proceedings of the 18th ACM Conference Companion on Computer Supported Cooperative Work \& Social Computing},
pages = {319–324},
numpages = {6},
location = {Vancouver, BC, Canada},
series = {CSCW'15 Companion}
}

@article{langan2004mental,
  title={Mental models, team mental models, and performance: Process, development, and future directions},
  author={Langan-Fox, Janice and Anglim, Jeromy and Wilson, John R},
  journal={Human Factors and Ergonomics in Manufacturing \& Service Industries},
  volume={14},
  number={4},
  pages={331--352},
  year={2004},
  publisher={Wiley Online Library}
}

@article{mathieu2000influence,
  title={The influence of shared mental models on team process and performance.},
  author={Mathieu, John E and Heffner, Tonia S and Goodwin, Gerald F and Salas, Eduardo and Cannon-Bowers, Janis A},
  journal={Journal of applied psychology},
  volume={85},
  number={2},
  pages={273},
  year={2000},
  publisher={American Psychological Association}
}

@inproceedings{Yi2008understandinginsights,
author = {Yi, Ji Soo and Kang, Youn-ah and Stasko, John T. and Jacko, Julie A.},
title = {Understanding and characterizing insights: how do people gain insights using information visualization?},
year = {2008},
isbn = {9781605580166},
publisher = {Association for Computing Machinery},
address = {New York, NY, USA},
url = {https://doi.org/10.1145/1377966.1377971},
doi = {10.1145/1377966.1377971},
booktitle = {Proceedings of the 2008 Workshop on BEyond Time and Errors: Novel EvaLuation Methods for Information Visualization},
articleno = {4},
numpages = {6},
location = {Florence, Italy},
series = {BELIV '08}
}

@article{olson2000distance,
  title={Distance matters},
  author={Olson, Gary M and Olson, Judith S},
  journal={Human--computer interaction},
  volume={15},
  number={2-3},
  pages={139--178},
  year={2000},
  publisher={Taylor \& Francis}
}

@inproceedings{Lee2022DistractingVideoConferencingPandemic,
author = {Lee, Minha and Park, Wonyoung and Lee, Sunok and Lee, Sangsu},
title = {Distracting Moments in Videoconferencing: A Look Back at the Pandemic Period},
year = {2022},
isbn = {9781450391573},
publisher = {Association for Computing Machinery},
address = {New York, NY, USA},
url = {https://doi.org/10.1145/3491102.3517545},
doi = {10.1145/3491102.3517545},
booktitle = {Proceedings of the 2022 CHI Conference on Human Factors in Computing Systems},
articleno = {141},
numpages = {21},
location = {New Orleans, LA, USA},
series = {CHI '22}
}

@article{fiore2016,
  title = {Technology as {{Teammate}}: {{Examining}} the {{Role}} of {{External Cognition}} in {{Support}} of {{Team Cognitive Processes}}},
  shorttitle = {Technology as {{Teammate}}},
  author = {Fiore, Stephen M. and Wiltshire, Travis J.},
  year = {2016},
  month = oct,
  journal = {Frontiers in Psychology},
  volume = {7},
  publisher = {Frontiers},
  issn = {1664-1078},
  doi = {10.3389/fpsyg.2016.01531},
  urldate = {2025-04-30},
  langid = {english}
}

@article{fiore2010,
  title = {Toward an {{Understanding}} of {{Macrocognition}} in {{Teams}}: {{Predicting Processes}} in {{Complex Collaborative Contexts}}},
  shorttitle = {Toward an {{Understanding}} of {{Macrocognition}} in {{Teams}}},
  author = {Fiore, Stephen M. and Rosen, Michael A. and {Smith-Jentsch}, Kimberly A. and Salas, Eduardo and Letsky, Michael and Warner, Norman},
  year = {2010},
  month = apr,
  journal = {Human Factors},
  volume = {52},
  number = {2},
  pages = {203--224},
  publisher = {SAGE Publications Inc},
  issn = {0018-7208},
  doi = {10.1177/0018720810369807},
  urldate = {2025-04-30},
  langid = {english}
}

@inproceedings{Martinez2011Collaid,
author = {Mart\'{\i}nez, Roberto and Collins, Anthony and Kay, Judy and Yacef, Kalina},
title = {Who Did What? Who Said That? Collaid: An Environment for Capturing Traces of Collaborative Learning at the Tabletop},
year = {2011},
isbn = {9781450308717},
publisher = {Association for Computing Machinery},
address = {New York, NY, USA},
url = {https://doi.org/10.1145/2076354.2076387},
doi = {10.1145/2076354.2076387},
booktitle = {Proceedings of the ACM International Conference on Interactive Tabletops and Surfaces},
pages = {172–181},
numpages = {10},
location = {Kobe, Japan},
series = {ITS '11}
}

@inproceedings{thompson2021ambidots,
  title={AmbiDots: An Ambient Interface to Mediate Casual Social Settings through Peripheral Interaction},
  author={Thompson, Edward and Potts, Dominic and Hardy, John and Porter, Barry and Houben, Steven},
  booktitle={Proceedings of the 2021 Australian Conference on Human-Computer Interaction (OzCHI’21)},
  year={2021}
  }

@article{weiser1991computer,
  title={The Computer for the 21 st Century},
  author={Weiser, Mark},
  journal={Scientific american},
  volume={265},
  number={3},
  pages={94--105},
  year={1991},
  publisher={JSTOR}
}

@inproceedings{holmquist1999token,
  title={Token-based access to digital information},
  author={Holmquist, Lars Erik and Redstr{\"o}m, Johan and Ljungstrand, Peter},
  booktitle={Handheld and Ubiquitous Computing: First International Symposium, HUC’99 Karlsruhe, Germany, September 27--29, 1999 Proceedings 1},
  pages={234--245},
  year={1999},
  organization={Springer}
}

@article{price2004letsgetohys,
  title={Let’s get physical: The learning benefits of interacting in digitally augmented physical spaces},
  author={Price, Sara and Rogers, Yvonne},
  journal={Computers \& Education},
  volume={43},
  number={1-2},
  pages={137--151},
  year={2004},
  publisher={Elsevier}
}

@article{erickson2000,
author = {Erickson, Thomas and Kellogg, Wendy A.},
title = {Social translucence: an approach to designing systems that support social processes},
year = {2000},
issue_date = {March 2000},
publisher = {Association for Computing Machinery},
address = {New York, NY, USA},
volume = {7},
number = {1},
issn = {1073-0516},
url = {https://doi.org/10.1145/344949.345004},
doi = {10.1145/344949.345004},
journal = {ACM Trans. Comput.-Hum. Interact.},
month = {mar},
pages = {59–83},
numpages = {25}
}

@inproceedings{smit2022sense,
author = {Smit, Doroth\'{e} and Hengeveld, Bart and Murer, Martin and Tscheligi, Manfred},
title = {Hybrid Design Tools for Participatory, Embodied Sensemaking: An Applied Framework},
year = {2022},
isbn = {9781450391474},
publisher = {Association for Computing Machinery},
address = {New York, NY, USA},
url = {https://doi.org/10.1145/3490149.3501332},
doi = {10.1145/3490149.3501332},
booktitle = {Proceedings of the Sixteenth International Conference on Tangible, Embedded, and Embodied Interaction},
articleno = {26},
numpages = {10},
location = {Daejeon, Republic of Korea},
series = {TEI '22}
}

@InProceedings{Messeguer2008,
author="Messeguer, Roc
and Dami{\'a}n-Reyes, Pedro
and Favela, Jesus
and Navarro, Leandro",
editor="Briggs, Robert O.
and Antunes, Pedro
and de Vreede, Gert-Jan
and Read, Aaron S.",
title="Context Awareness and Uncertainty in Collocated Collaborative Systems",
booktitle="Groupware: Design, Implementation, and Use",
year="2008",
publisher="Springer Berlin Heidelberg",
address="Berlin, Heidelberg",
pages="41--56",
isbn="978-3-540-92831-7"
}

@inproceedings{Klinkhammer2018tabletop,
author = {Klinkhammer, Daniel and Mateescu, Magdalena and Zahn, Carmen and Reiterer, Harald},
title = {Mine, Yours, Ours: Coordination through Workspace Arrangements and Territoriality in Tabletop Interaction},
year = {2018},
isbn = {9781450365949},
publisher = {Association for Computing Machinery},
address = {New York, NY, USA},
url = {https://doi.org/10.1145/3282894.3282902},
doi = {10.1145/3282894.3282902},
booktitle = {Proceedings of the 17th International Conference on Mobile and Ubiquitous Multimedia},
pages = {171–182},
numpages = {12},
location = {Cairo, Egypt},
series = {MUM '18}
}

@inproceedings{Brudy2019CrossDevice,
author = {Brudy, Frederik and Holz, Christian and R\"{a}dle, Roman and Wu, Chi-Jui and Houben, Steven and Klokmose, Clemens Nylandsted and Marquardt, Nicolai},
title = {Cross-Device Taxonomy: Survey, Opportunities and Challenges of Interactions Spanning Across Multiple Devices},
year = {2019},
isbn = {9781450359702},
publisher = {Association for Computing Machinery},
address = {New York, NY, USA},
url = {https://doi.org/10.1145/3290605.3300792},
doi = {10.1145/3290605.3300792},
booktitle = {Proceedings of the 2019 CHI Conference on Human Factors in Computing Systems},
pages = {1–28},
numpages = {28},
location = {Glasgow, Scotland Uk},
series = {CHI '19}
}

@article{hutchins1985direct,
  title={Direct manipulation interfaces},
  author={Hutchins, Edwin L and Hollan, James D and Norman, Donald A},
  journal={Human--computer interaction},
  volume={1},
  number={4},
  pages={311--338},
  year={1985},
  publisher={Taylor \& Francis}
}

@article{shneiderman1983direct,
  title={Direct manipulation: A step beyond programming languages},
  author={Shneiderman, Ben},
  journal={Computer},
  volume={16},
  number={08},
  pages={57--69},
  year={1983},
  publisher={IEEE Computer Society}
}

@article{Weiser1993,
author = {Weiser, Mark},
title = {Some Computer Science Issues in Ubiquitous Computing},
year = {1993},
issue_date = {July 1993},
publisher = {Association for Computing Machinery},
address = {New York, NY, USA},
volume = {36},
number = {7},
issn = {0001-0782},
url = {https://doi.org/10.1145/159544.159617},
doi = {10.1145/159544.159617},
journal = {Commun. ACM},
month = {jul},
pages = {75–84},
numpages = {10}
}

@article{Clayphan2014scriptstorm,
author = {Clayphan, Andrew and Kay, Judy and Weinberger, Armin},
title = {ScriptStorm: Scripting to Enhance Tabletop Brainstorming},
year = {2014},
issue_date = {August    2014},
publisher = {Springer-Verlag},
address = {Berlin, Heidelberg},
volume = {18},
number = {6},
issn = {1617-4909},
url = {https://doi.org/10.1007/s00779-013-0746-z},
doi = {10.1007/s00779-013-0746-z},
journal = {Personal Ubiquitous Comput.},
month = {aug},
pages = {1433–1453},
numpages = {21}
}

@inproceedings{Clayphan2011Firestorm,
author = {Clayphan, Andrew and Collins, Anthony and Ackad, Christopher and Kummerfeld, Bob and Kay, Judy},
title = {Firestorm: A Brainstorming Application for Collaborative Group Work at Tabletops},
year = {2011},
isbn = {9781450308717},
publisher = {Association for Computing Machinery},
address = {New York, NY, USA},
url = {https://doi.org/10.1145/2076354.2076386},
doi = {10.1145/2076354.2076386},
booktitle = {Proceedings of the ACM International Conference on Interactive Tabletops and Surfaces},
pages = {162–171},
numpages = {10},
location = {Kobe, Japan},
series = {ITS '11}
}

@article{hunter2008wordplay,
  title={WordPlay: A table-top interface for collaborative brainstorming and decision making},
  author={Hunter, Seth and Maes, Pattie},
  journal={Proceedings of IEEE Tabletops and Interactive Surfaces},
  pages={2--5},
  year={2008},
  publisher={Citeseer}
}

@article{divesta1973,
  title = {Listening and Note Taking: {{II}}. {{Immediate}} and Delayed Recall as Functions of Variations in Thematic Continuity, Note Taking, and Length of Listening-Review Intervals},
  shorttitle = {Listening and Note Taking},
  author = {Di Vesta, Francis J. and Gray, G. Susan},
  year = {1973},
  journal = {Journal of Educational Psychology},
  volume = {64},
  number = {3},
  pages = {278--287},
  publisher = {American Psychological Association},
  address = {US},
  issn = {1939-2176},
  doi = {10.1037/h0034589}
}

@article{mueller2014c,
  title = {The {{Pen Is Mightier Than}} the {{Keyboard}}: {{Advantages}} of {{Longhand Over Laptop Note Taking}}},
  shorttitle = {The {{Pen Is Mightier Than}} the {{Keyboard}}},
  author = {Mueller, Pam A. and Oppenheimer, Daniel M.},
  year = {2014},
  month = jun,
  journal = {Psychological Science},
  volume = {25},
  number = {6},
  pages = {1159--1168},
  publisher = {SAGE Publications Inc},
  issn = {0956-7976},
  doi = {10.1177/0956797614524581},
  urldate = {2025-05-05},
  langid = {english}
}

@inproceedings{Wells2020CollabAR,
author = {Wells, Thomas and Houben, Steven},
title = {CollabAR – Investigating the Mediating Role of Mobile AR Interfaces on Co-Located Group Collaboration},
year = {2020},
isbn = {9781450367080},
publisher = {Association for Computing Machinery},
address = {New York, NY, USA},
url = {https://doi.org/10.1145/3313831.3376541},
doi = {10.1145/3313831.3376541},
booktitle = {Proceedings of the 2020 CHI Conference on Human Factors in Computing Systems},
pages = {1–13},
numpages = {13},
location = {Honolulu, HI, USA},
series = {CHI '20}
}

@inproceedings{Reactable2007,
author = {Jord\`{a}, Sergi and Geiger, G\"{u}nter and Alonso, Marcos and Kaltenbrunner, Martin},
title = {The ReacTable: Exploring the Synergy between Live Music Performance and Tabletop Tangible Interfaces},
year = {2007},
isbn = {9781595936196},
publisher = {Association for Computing Machinery},
address = {New York, NY, USA},
url = {https://doi.org/10.1145/1226969.1226998},
doi = {10.1145/1226969.1226998},
booktitle = {Proceedings of the 1st International Conference on Tangible and Embedded Interaction},
pages = {139–146},
numpages = {8},
location = {Baton Rouge, Louisiana},
series = {TEI '07}
}

@inproceedings{Brudy2018overview,
author = {Brudy, Frederik and Budiman, Joshua Kevin and Houben, Steven and Marquardt, Nicolai},
title = {Investigating the Role of an Overview Device in Multi-Device Collaboration},
year = {2018},
isbn = {9781450356206},
publisher = {Association for Computing Machinery},
address = {New York, NY, USA},
url = {https://doi.org/10.1145/3173574.3173874},
doi = {10.1145/3173574.3173874},
booktitle = {Proceedings of the 2018 CHI Conference on Human Factors in Computing Systems},
pages = {1–13},
numpages = {13},
location = {Montreal QC, Canada},
series = {CHI '18}
}

@inproceedings{brudy2016curationspace,
  title={Curationspace: Cross-device content curation using instrumental interaction},
  author={Brudy, Frederik and Houben, Steven and Marquardt, Nicolai and Rogers, Yvonne},
  booktitle={Proceedings of the 2016 ACM International Conference on Interactive Surfaces and Spaces},
  publisher = {Association for Computing Machinery},
  pages={159--168},
  year={2016}
}

@article{wallace2011investigating,
  title={Investigating the role of a large, shared display in multi-display environments},
  author={Wallace, James R and Scott, Stacey D and Lai, Eugene and Jajalla, Deon},
  journal={Computer Supported Cooperative Work (CSCW)},
  volume={20},
  pages={529--561},
  year={2011},
  publisher={Springer}
}

@article{kruger2004roles,
  title={Roles of orientation in tabletop collaboration: Comprehension, coordination and communication},
  author={Kruger, Russell and Carpendale, Sheelagh and Scott, Stacey D and Greenberg, Saul},
  journal={Computer Supported Cooperative Work (CSCW)},
  volume={13},
  pages={501--537},
  year={2004},
  publisher={Springer}
}

@article{reusser2015co,
  title={Co-constructivism in educational theory and practice},
  author={Reusser, Kurt and Pauli, Christine and Wright, James D},
  year={2015},
  publisher={Elsevier}
}

@inproceedings{vogt2011co,
  title={Co-located collaborative sensemaking on a large high-resolution display with multiple input devices},
  author={Vogt, Katherine and Bradel, Lauren and Andrews, Christopher and North, Chris and Endert, Alex and Hutchings, Duke},
  booktitle={Human-Computer Interaction--INTERACT 2011: 13th IFIP TC 13 International Conference, Lisbon, Portugal, September 5-9, 2011, Proceedings, Part II 13},
  pages={589--604},
  year={2011},
  organization={Springer}
}

@article{van2015communication,
  title={Communication, concepts and grounding},
  author={van der Velde, Frank},
  journal={Neural networks},
  volume={62},
  pages={112--117},
  year={2015},
  publisher={Elsevier}
}

@inproceedings{Shi2017talktrawallCreativeCollabVisualStimuli,
author = {Shi, Yang and Wang, Yang and Qi, Ye and Chen, John and Xu, Xiaoyao and Ma, Kwan-Liu},
title = {IdeaWall: Improving Creative Collaboration through Combinatorial Visual Stimuli},
year = {2017},
isbn = {9781450343350},
publisher = {Association for Computing Machinery},
address = {New York, NY, USA},
url = {https://doi.org/10.1145/2998181.2998208},
doi = {10.1145/2998181.2998208},
booktitle = {Proceedings of the 2017 ACM Conference on Computer Supported Cooperative Work and Social Computing},
pages = {594–603},
numpages = {10},
location = {Portland, Oregon, USA},
series = {CSCW '17}
}

@inproceedings{chandra1029TalkTracesRealtimeCaptureandVisofVerbalContentinMeetings,
author = {Chandrasegaran, Senthil and Bryan, Chris and Shidara, Hidekazu and Chuang, Tung-Yen and Ma, Kwan-Liu},
title = {TalkTraces: Real-Time Capture and Visualization of Verbal Content in Meetings},
year = {2019},
isbn = {9781450359702},
publisher = {Association for Computing Machinery},
address = {New York, NY, USA},
url = {https://doi.org/10.1145/3290605.3300807},
doi = {10.1145/3290605.3300807},
booktitle = {Proceedings of the 2019 CHI Conference on Human Factors in Computing Systems},
pages = {1–14},
numpages = {14},
location = {Glasgow, Scotland Uk},
series = {CHI '19}
}

@inproceedings{gronbaek2020,
author = {Gr\o{}nb\ae{}k, Jens Emil and Rasmussen, Majken Kirkegaard and Halskov, Kim and Petersen, Marianne Graves},
title = {KirigamiTable: Designing for Proxemic Transitions with a Shape-Changing Tabletop},
year = {2020},
isbn = {9781450367080},
publisher = {Association for Computing Machinery},
address = {New York, NY, USA},
url = {https://doi.org/10.1145/3313831.3376834},
doi = {10.1145/3313831.3376834},
booktitle = {Proceedings of the 2020 CHI Conference on Human Factors in Computing Systems},
pages = {1–15},
numpages = {15},
location = {Honolulu, HI, USA},
series = {CHI '20}
}

@inproceedings{jaasma2017,
author = {Jaasma, Phil\'{e}monne and Wolters, Evert and Frens, Joep and Hummels, Caroline and Trotto, Ambra},
title = {[X]Changing Perspectives: An Interactive System for Participatory Sensemaking},
year = {2017},
isbn = {9781450349222},
publisher = {Association for Computing Machinery},
address = {New York, NY, USA},
url = {https://doi.org/10.1145/3064663.3064796},
doi = {10.1145/3064663.3064796},
booktitle = {Proceedings of the 2017 Conference on Designing Interactive Systems},
pages = {259–269},
numpages = {11},
location = {Edinburgh, United Kingdom},
series = {DIS '17}
}

@inproceedings{serim2019,
  title = {Explicating "{{Implicit Interaction}}": {{An Examination}} of the {{Concept}} and {{Challenges}} for {{Research}}},
  shorttitle = {Explicating "{{Implicit Interaction}}"},
  booktitle = {Proceedings of the 2019 {{CHI Conference}} on {{Human Factors}} in {{Computing Systems}}},
  author = {Serim, Bar{\i}{\c s} and Jacucci, Giulio},
  year = {2019},
  month = may,
  series = {{{CHI}} '19},
  pages = {1--16},
  publisher = {Association for Computing Machinery},
  address = {New York, NY, USA},
  doi = {10.1145/3290605.3300647},
  urldate = {2025-05-06},
  isbn = {978-1-4503-5970-2}
}

@article{martinez2019collocated,
  title={Collocated collaboration analytics: Principles and dilemmas for mining multimodal interaction data},
  author={Martinez-Maldonado, Roberto and Kay, Judy and Buckingham Shum, Simon and Yacef, Kalina},
  journal={Human--Computer Interaction},
  volume={34},
  number={1},
  pages={1--50},
  year={2019},
  publisher={Taylor \& Francis}
}

@article{hevey2018network,
  title={Network analysis: a brief overview and tutorial},
  author={Hevey, David},
  journal={Health Psychology and Behavioral Medicine},
  volume={6},
  number={1},
  pages={301--328},
  year={2018},
  publisher={Taylor \& Francis}
}


\appendix


\end{document}